\documentclass[a4paper,fleqn]{cas-dc} 

\usepackage{hyperref}       
\usepackage{url}            
\usepackage[utf8]{inputenc} 
\usepackage[numbers]{natbib}
\usepackage{subfigure}

\def\tsc#1{\csdef{#1}{\textsc{\lowercase{#1}}\xspace}}
\tsc{WGM}
\tsc{QE}

\begin{document}
\let\WriteBookmarks\relax
\def\floatpagepagefraction{1}
\def\textpagefraction{.001}

\shorttitle{}    

\shortauthors{}  

\title [mode = title]{Measurements on Time Resolution of BGO, PWO and BSO Crystals}  


%

\author[3,4,5]{Zhiyu Zhao}[type=editor]

\author[1,2]{Dejing Du}

\author[1,2]{Yong Liu}
\cormark[1]
\cortext[1]{Corresponding author}
\ead{liuyong@ihep.ac.cn}

\author[4,5,3]{Jiyuan Chen}

\author[6,7]{Junfeng Chen}

\author[1,2]{Fangyi Guo}

\author[3,4,5]{Shu Li}

\author[1,2]{Baohua Qi}

\affiliation[1]{organization={Institute of High Energy Physics, Chinese Academy of Sciences},
            addressline={19B Yuquan Road}, 
            city={Beijing},
            postcode={100049}, 
            country={China}}

\affiliation[2]{organization={University of Chinese Academy of Sciences},
            addressline={19A Yuquan Road}, 
            city={Beijing},
            postcode={100049}, 
            country={China}}

\affiliation[3]{organization={Tsung-Dao Lee Institute, Shanghai Jiao Tong University},
            addressline={1 Lisuo Road}, 
            city={Shanghai},
            postcode={201210}, 
            country={China}}

\affiliation[4]{organization={Institute of Nuclear and Particle Physics, School of Physics and Astronomy},
            addressline={800 Dongchuan Road}, 
            city={Shanghai},
            postcode={200240}, 
            country={China}}

\affiliation[5]{organization={Key Laboratory for Particle Astrophysics and Cosmology (MOE), Shanghai Key Laboratory for Particle Physics and Cosmology (SKLPPC), Shanghai Jiao Tong University},
            addressline={800 Dongchuan Road}, 
            city={Shanghai},
            postcode={200240}, 
            country={China}}

\affiliation[6]{organization={Center of Materials Science and Optoelectronics Engineering, University of Chinese Academy of Science},
            addressline={19A Yuquan Road}, 
            city={Beijing},
            postcode={100049}, 
            country={China}}
            
\affiliation[7]{organization={Shanghai Institute of Ceramics, Chinese Academy of Sciences},
            addressline={1295 Dingxi Road}, 
            city={Shanghai},
            postcode={201899}, 
            country={China}}

\begin{abstract}
A high-granularity crystal calorimeter (HGCCAL) has been proposed for the future Circular Electron Positron Collider (CEPC). This study investigates the time resolution of various crystal–Silicon Photomultiplier (SiPM) detection units for HGCCAL, focusing on Bismuth Germanate (BGO), Lead Tungstate (PWO), and Bismuth Silicon Oxide (BSO) crystals. Beam tests were conducted using 10 GeV pions at CERN and 5 GeV electrons at DESY, enabling systematic comparisons of timing performance under both minimum ionizing particle (MIP) signals and electromagnetic (EM) showers. Three timing methods—constant fraction timing (CFT) with sampled points, linear fitting, and exponential fitting—were evaluated, with an exponential fit combined with a 10\% constant fraction providing the best time resolution.

Measurements of crystal units with different dimensions revealed that both scintillation light yield and signal rise time influence timing performance. Among similarly sized crystals, PWO exhibited the best time resolution due to its fast signal rise time, while BGO and BSO demonstrated comparable timing performance. For long BGO bars (40 cm and 60 cm), the time resolution remained uniform along their length, achieving approximately 0.75 ns and 0.95 ns for MIP signals. Under intense EM showers, both bars reached a timing resolution of approximately 200 ps at high amplitudes. And the presence of upstream pre-shower layers can introduce additional timing fluctuations at similar amplitudes.

\end{abstract}




\begin{keywords}
Crystal scintillator \sep BGO \sep PWO \sep BSO \sep Time Resolution \sep SiPM \sep Electromagnetic Shower \sep Calorimeter \sep Higgs Factory \sep CEPC \sep Beam Test
\end{keywords}

\maketitle

\section{Introduction}
\label{sec:intro}

The Circular Electron Positron Collider (CEPC)~\cite{CEPCStudyGroup:2023quu, CEPCCDR-2} is a proposed next-generation electron-positron collider designed to function as a Higgs factory, aiming for precise measurements of the Higgs boson, W/Z bosons, top quark, and potential new physics beyond the Standard Model. To achieve high jet energy resolution, a high-granularity calorimeter based on the particle-flow algorithm (PFA) has emerged as a leading option. Crystal calorimeters have demonstrated exceptional performance in experimental particle physics, with notable examples including the CMS Electromagnetic Calorimeter (ECAL)~\cite{CERN-LHCC-97-033}, which played a crucial role in the discovery of the Higgs boson~\cite{CMS:2012qbp} at the Large Hadron Collider (LHC); the BGO electromagnetic calorimeter~\cite{SUMNER1988252} in the L3 experiment~\cite{ADEVA199035} at CERN's Large Electron-Positron Collider (LEP)~\cite{Myers:226776}, which contributed significantly to precision electroweak measurements; and ECALs employed in flavor physics experiments such as BaBar~\cite{BaBar:1995bns}, Belle II~\cite{Belle:1995pqe, abe2010belleiitechnicaldesign}, and BESIII~\cite{ABLIKIM2010345}.

A novel approach, the High-Granularity Crystal Calorimeter (HGCCAL)~\cite{Liu_2020, instruments6030040}, has been proposed to achieve excellent electromagnetic energy resolution while maintaining compatibility with the PFA~\cite{THOMSON200925} for CEPC. HGCCAL features a homogeneous structure with Silicon Photon Multipliers (SiPMs) as the baseline readout sensors, and dedicated research and development (R\&D) efforts have been initiated to advance this concept.

Among the various high-density scintillating crystals evaluated for HGCCAL, Bismuth Germanate (BGO)~\cite{JI2014143} has been selected as the baseline option due to its high intrinsic light yield, making it highly efficient in detecting electromagnetic particles. Additionally, the technology for growing long BGO bars has been well-established and demonstrated. In addition to BGO, Lead Tungstate (PWO)~\cite{BACCARO199866} and Bismuth Silicon Oxide (BSO)~\cite{ISHII2002201} crystals are also being actively studied due to their distinct advantages. PWO crystals offer a significantly fast scintillation response, improving timing performance, though at the cost of a lower light yield compared to BGO. Meanwhile, BSO, as a relatively new scintillator material, provides a moderate light yield, a faster scintillation decay time, and is more cost-effective than BGO. The choice of crystal material directly impacts calorimeter performance, particularly in energy and time resolution, making the evaluation of the basic crystal detection unit essential for optimizing HGCCAL.

Time resolution is a fundamental parameter in detector performance, determining the precision with which the detector measures the arrival time of a particle. High time resolution is crucial for distinguishing closely spaced events, reducing background noise, and enhancing particle identification and energy measurement accuracy. In the context of the CEPC and HGCCAL, superior time resolution improves the reconstruction of electromagnetic showers, leading to enhanced energy measurement precision and particle identification capabilities. Time resolution is influenced by various detector components, including the intrinsic properties of scintillating crystals, the performance of SiPMs, and the electronic readout system.

This study evaluates the time resolution of crystal-SiPM detection units using BGO, PWO, and BSO crystals of different dimensions. Through a series of beam tests with both minimum ionizing particles (MIPs) and high-energy electrons, we investigate factors affecting time resolution, including crystal dimensions, scintillation properties, and electronics. Specifically, we examine the effectiveness of different timing methods, the uniformity of time response along long BGO crystal bars, and the impact of electromagnetic shower development on timing performance. The results provide valuable insights for optimizing crystal calorimeter designs for CEPC.

This study focuses on evaluating the time resolution of crystal-SiPM detection units using BGO, PWO, and BSO crystals of different dimensions. Through a series of beam tests involving both minimum ionizing particles (MIPs) and high-energy electrons, we have investigated the factors influencing time resolution, including crystal dimensions, scintillation properties, and electronics. Specifically, we examine the overall effectiveness of different timing methods, the uniformity of time response along long BGO crystal bars, and the impact of electromagnetic shower development on timing performance. The results provide valuable insights into the optimization of crystal calorimeter designs for CEPC.

\section{Experiment description}
\label{sec:ExpSetup}

The time resolution of various crystal scintillators was measured and compared through beam tests. Key factors affecting the MIP time resolution of individual crystals, such as crystal dimensions, timing techniques, and the intrinsic time resolution of the electronics, were investigated. Additionally, the time resolution of crystal units under electromagnetic showers was examined.

\subsection{Crystal samples}

\begin{figure}[h]
    \centering  
    \subfigure[]{
    \includegraphics[width=0.4\textwidth]{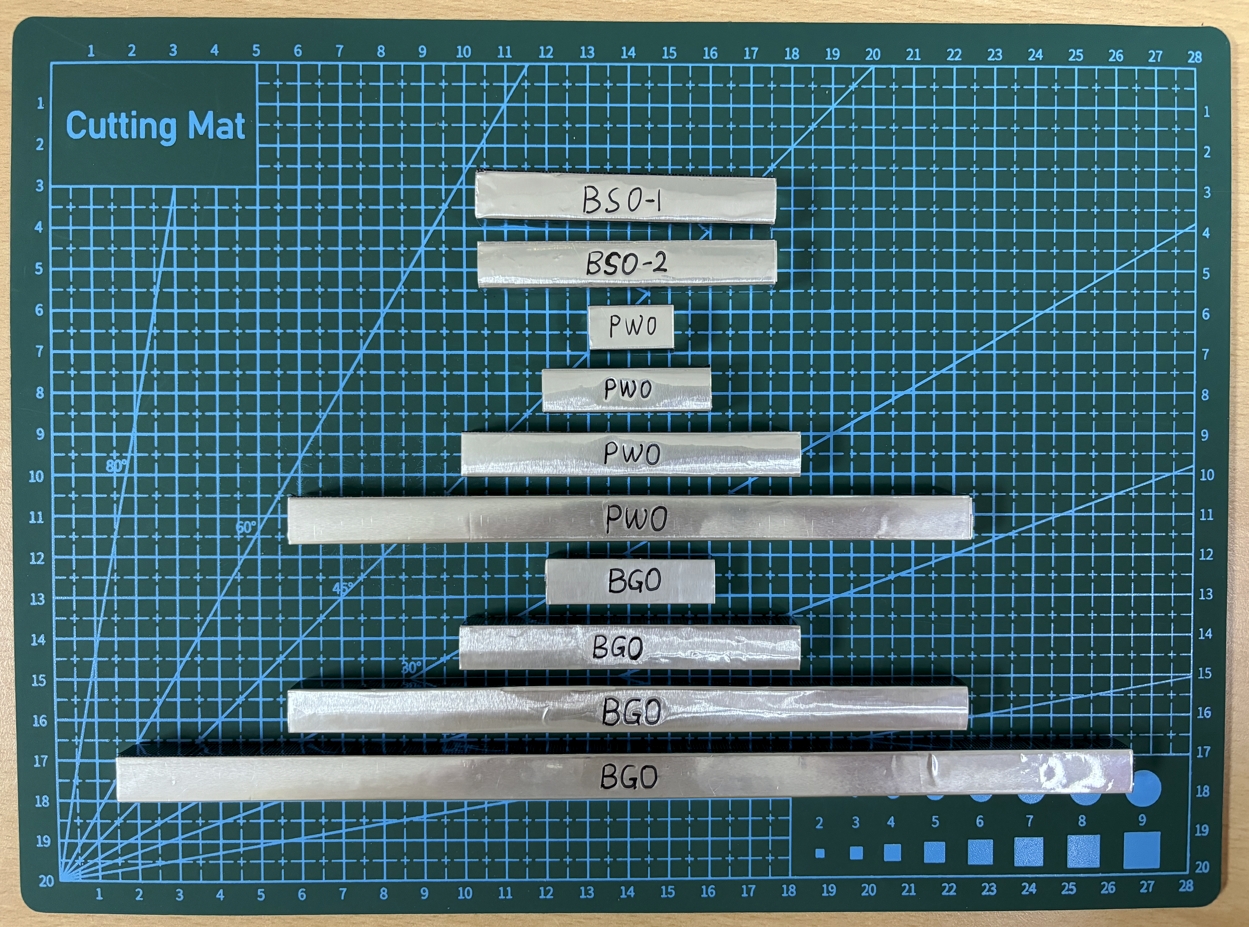}s}
    \subfigure[]{
    \includegraphics[width=0.4\textwidth]{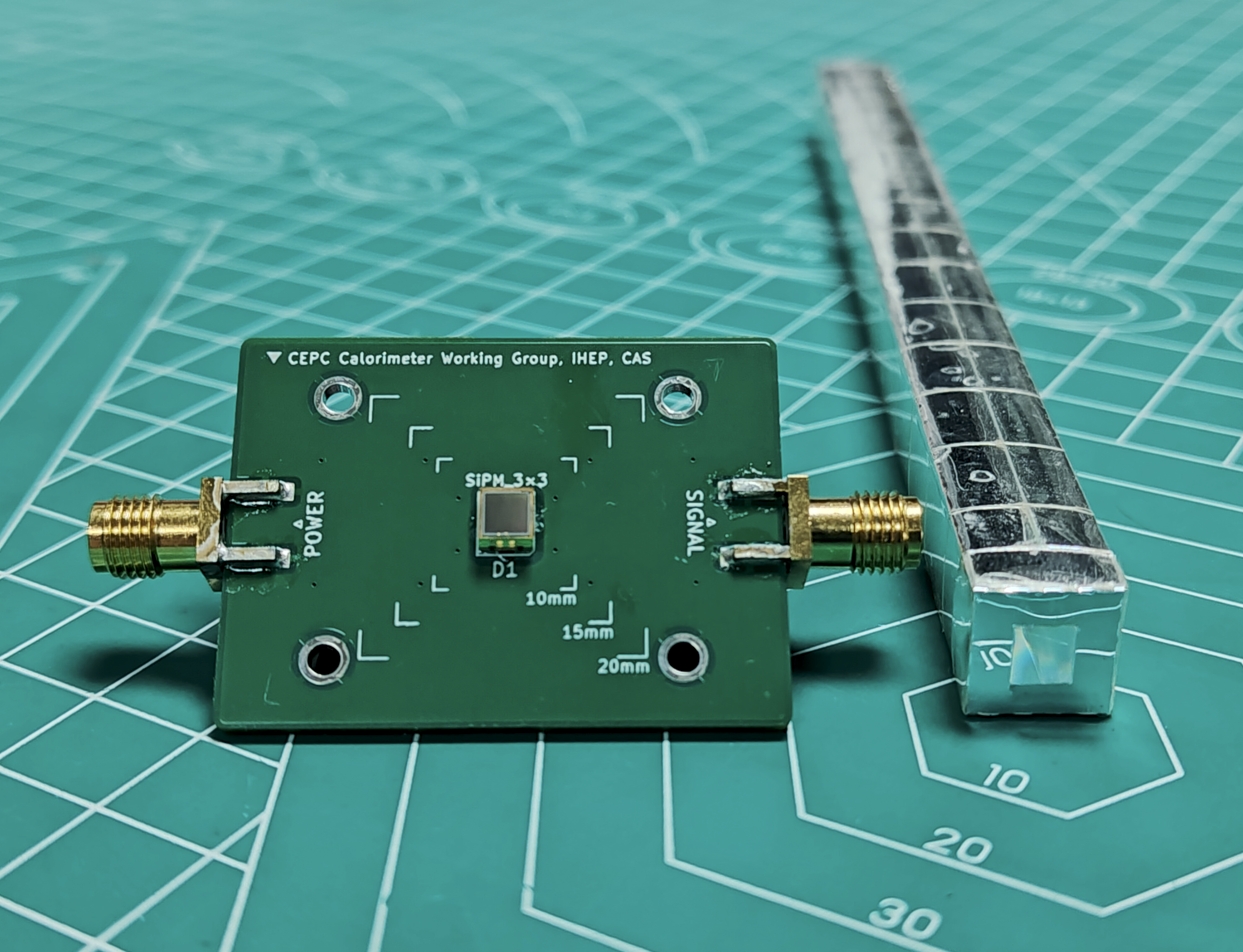}}
    \caption{\label{fig:BGO}~(a) BGO, PWO and BSO crystal scintillator with various sizes, wrapped in ESR and aluminum reflective films. (b) The basic Crystal-SiPM detection unit.}
\end{figure}

\begin{table*}[width=.9\linewidth,h]
\centering
\fontsize{8}{12}\selectfont
\caption{\label{tab:Crystals}~The tested crystal samples, along with their quantities and dimensions.}
    \begin{tabular}{ccc}
        \toprule
        \makecell[c]{Crystal} &\makecell[c]{Quantity} &\makecell[c]{Dimension (cm$^3$)} \\ 
        \midrule
        BGO  & 6 & 1$\times$1$\times$4, 1$\times$1$\times$8, 1$\times$1$\times$16, 1$\times$1$\times$24, 1$\times$1$\times$40, 1.5$\times$1.5$\times$60  \\
        PWO & 4 & 1$\times$1$\times$2, 1$\times$1$\times$4, 1$\times$1$\times$8, 1$\times$1$\times$16 \\
        BSO  & 1 & 1$\times$1$\times$7\\
        \bottomrule
    \end{tabular}
\end{table*}

The crystal samples measured are listed in Table~\ref{tab:Crystals}. Three types of crystals produced by SIC-CAS were tested: BGO, PWO, and BSO. Among these, the BGO crystal exhibits the highest intrinsic light yield but the slowest scintillation decay time, while the PWO crystal has the fastest scintillation decay time but the lowest light yield. The BSO crystal demonstrates moderate performance. As light yield and scintillation decay time are the two primary factors influencing time resolution, the analysis of the crystals' time resolution will focus on these two aspects.

The crystal lengths ranged from 2 cm to 60 cm. Except for a single 60 cm-long BGO crystal with a cross-sectional area of 1.5$\times$1.5 cm$^2$, all other crystals had a cross-sectional area of 1$\times$1 cm$^2$. Each crystal was wrapped in a 65 $\mu$m-thick ESR reflective film, further covered by a 50 $\mu$m-thick aluminum film. The films at both ends were left open to allow optical coupling with SiPMs (HAMAMATSU S13360-6025PE~\cite{S13360-6025PE}) for scintillation light collection. The coupling was achieved through air, with the window size approximately matching the SiPM’s external dimensions.


\subsection{Setup of beam test}

\begin{table}[width=.9\linewidth,h]
\centering
\caption{\label{tab:Beams}~Beam configurations.}
    \begin{tabular}{cccc}
        \toprule
        \makecell[c]{Particle} &\makecell[c]{Energy (GeV)} &\makecell[c]{Usage} &\makecell[c]{Station}\\ 
        \midrule
        $\pi^-$  & 10 & MIP timing & CERN PS-T9\\
        $e^-$  & 5 & EM timing & DESY TB-22\\
        \bottomrule
    \end{tabular}
\end{table}

As summarized in Table~\ref{tab:Beams}, charged pions from the CERN PS-T9 beamline~\cite{Parozzi:2024pfk} and electrons from the TB-22 beamline at DESY~\cite{DIENER2019265} were used in the experiment. A 10 GeV pion traversing a single crystal behaves as MIP without inducing showers. The energy deposited in the crystal is approximately proportional to the path length, making it well-suited for studying light yield and time resolution per unit deposited energy. Additionally, 5 GeV electron beams were employed to investigate the time resolution of the crystals across different energy levels and stages of electromagnetic (EM) shower development.

\begin{figure}[h]
    \centering  
    \subfigure[]{
    \includegraphics[width=0.45\textwidth]{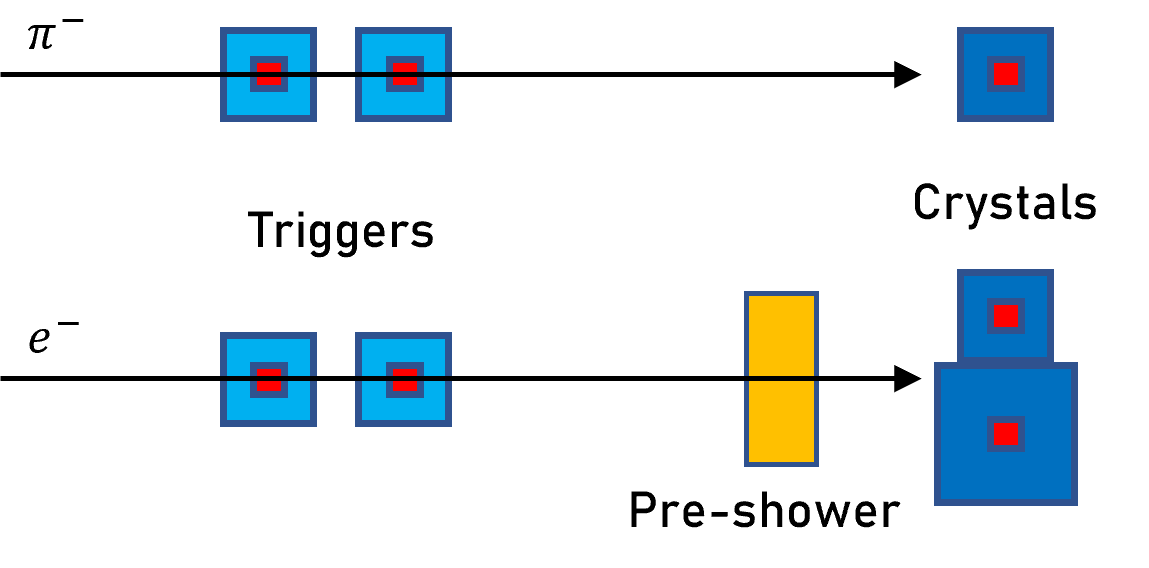}}
    \subfigure[]{
    \includegraphics[width=0.4\textwidth]{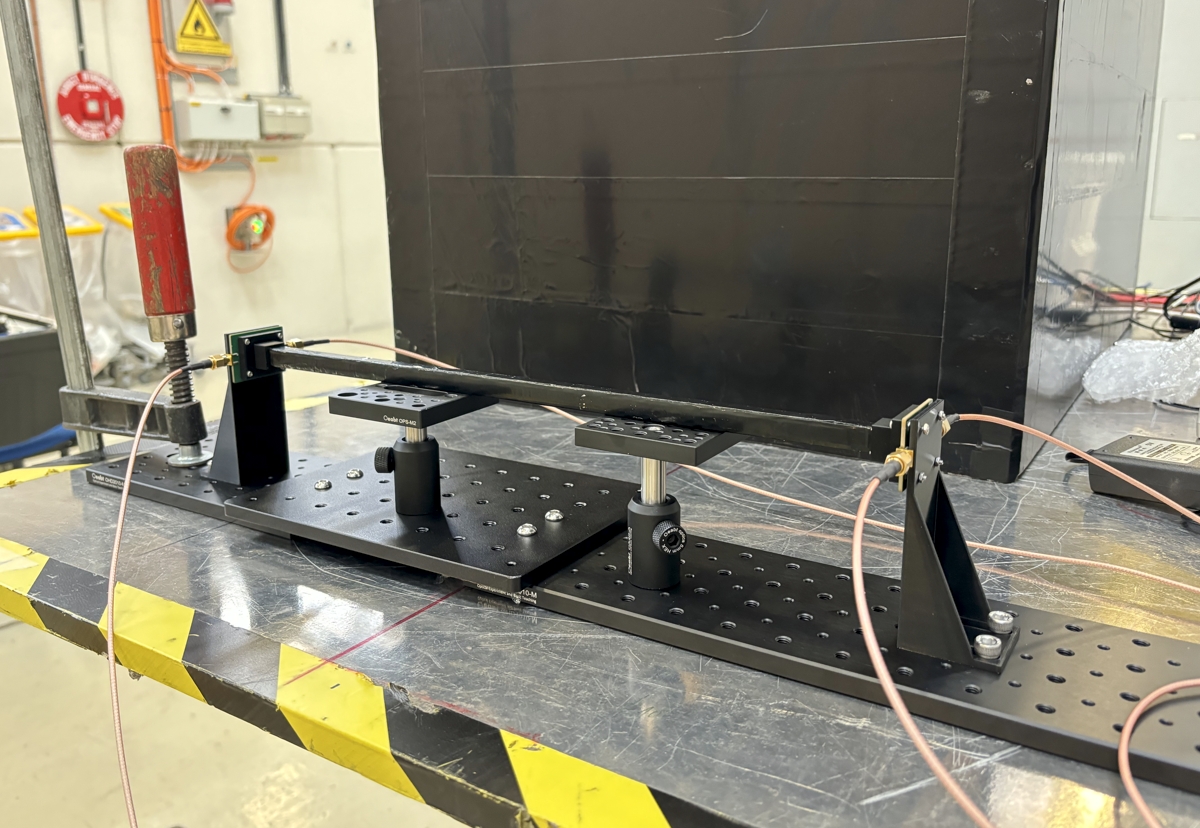}}
    \subfigure[]{
    \includegraphics[width=0.4\textwidth]{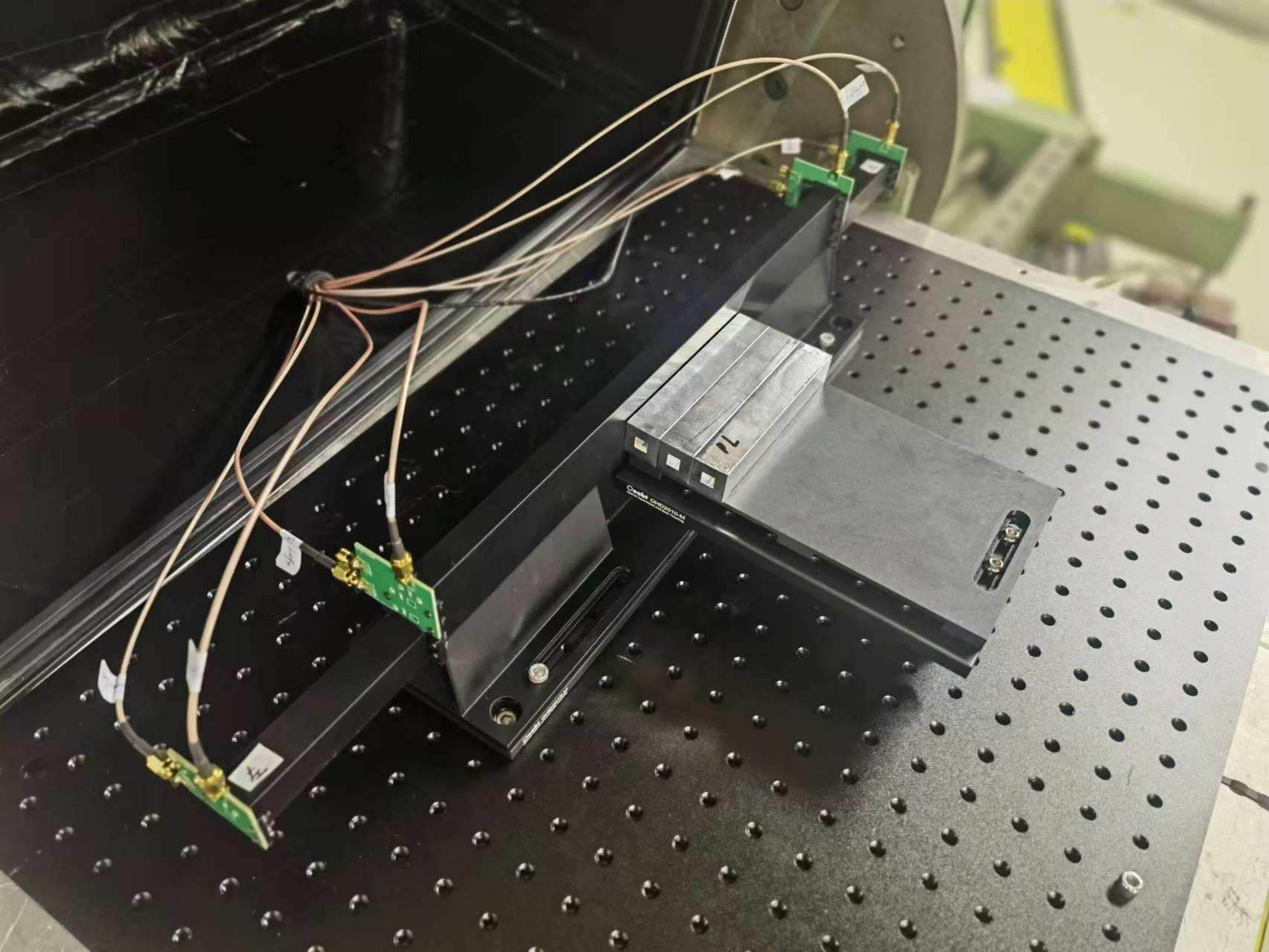}}
    \caption{\label{fig:BeamSetup}~Setup of beam test. Two 1cm$^3$ plastic scintillators coupled with SiPMs were placed upstream of the test crystal to provide trigger signals. (b) In $\pi^-$ beam test, particles passing the trigger hit the test crystal along the vertical direction of its length. (c) In $e^-$ beam test, a BGO pre-shower was placed upstream of the test crystals to capture the electromagnetic shower leakage from the pre-shower.}
\end{figure}

The beam test setup is illustrated in Figure~\ref{fig:BeamSetup}. The test crystal was positioned on a movable platform, aligned perpendicular to the beam direction, and enclosed in a light-tight black box (or curtain) to shield it from ambient light. Upstream of the crystal, two plastic scintillator units, each with a volume of 1 cm$^3$ and read out by SiPMs, provided the trigger signal.

For the MIP timing measurements with pions, the beam particles exited the beamline, triggered the plastic scintillators, and directly impacted the test crystals. In the electron beam tests, a pre-shower layer composed of BGO with varying thickness was placed between the trigger system and the test crystal. This setup allowed the investigation of the time resolution of the crystal bars at different stages of shower development and under various energy conditions.

\subsection{Time resolution of electronics}

\begin{figure}[h]
    \centering  
    \subfigure[]{
    \includegraphics[width=0.45\textwidth]{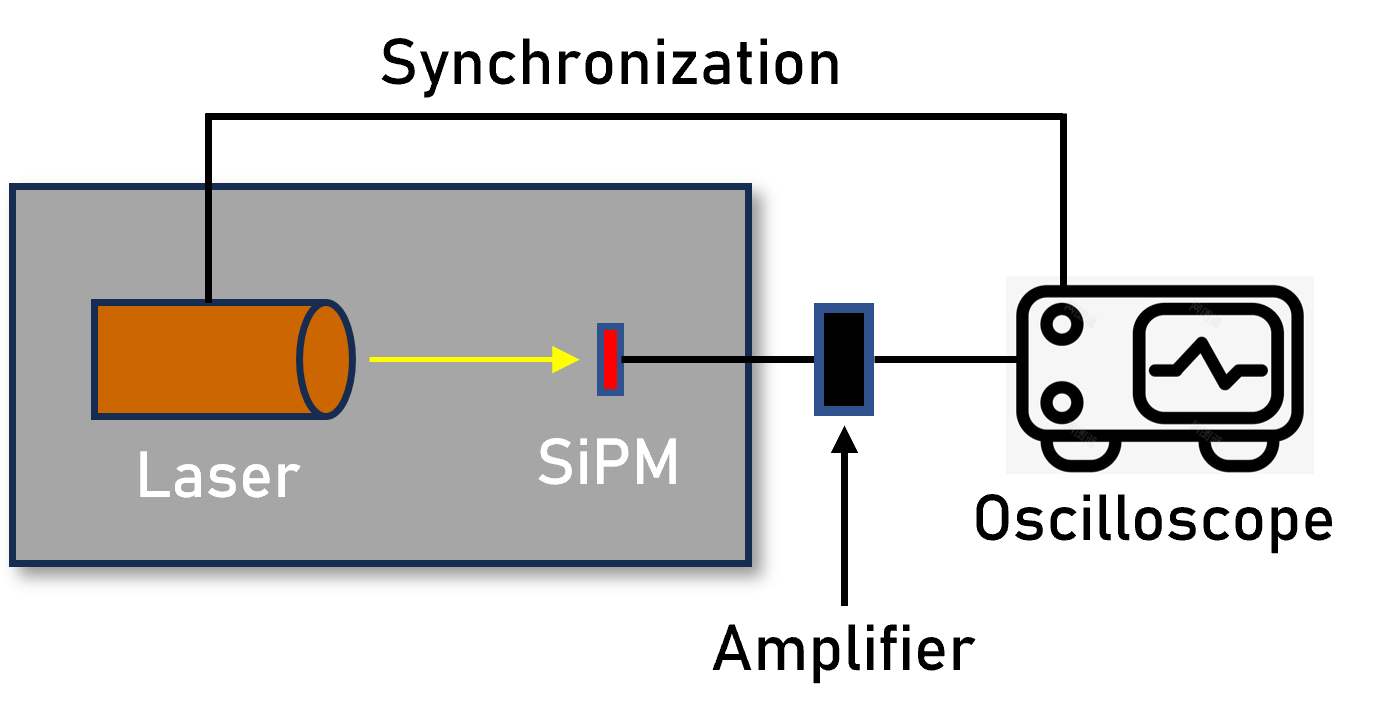}} 
    \subfigure[]{
    \includegraphics[width=0.45\textwidth]{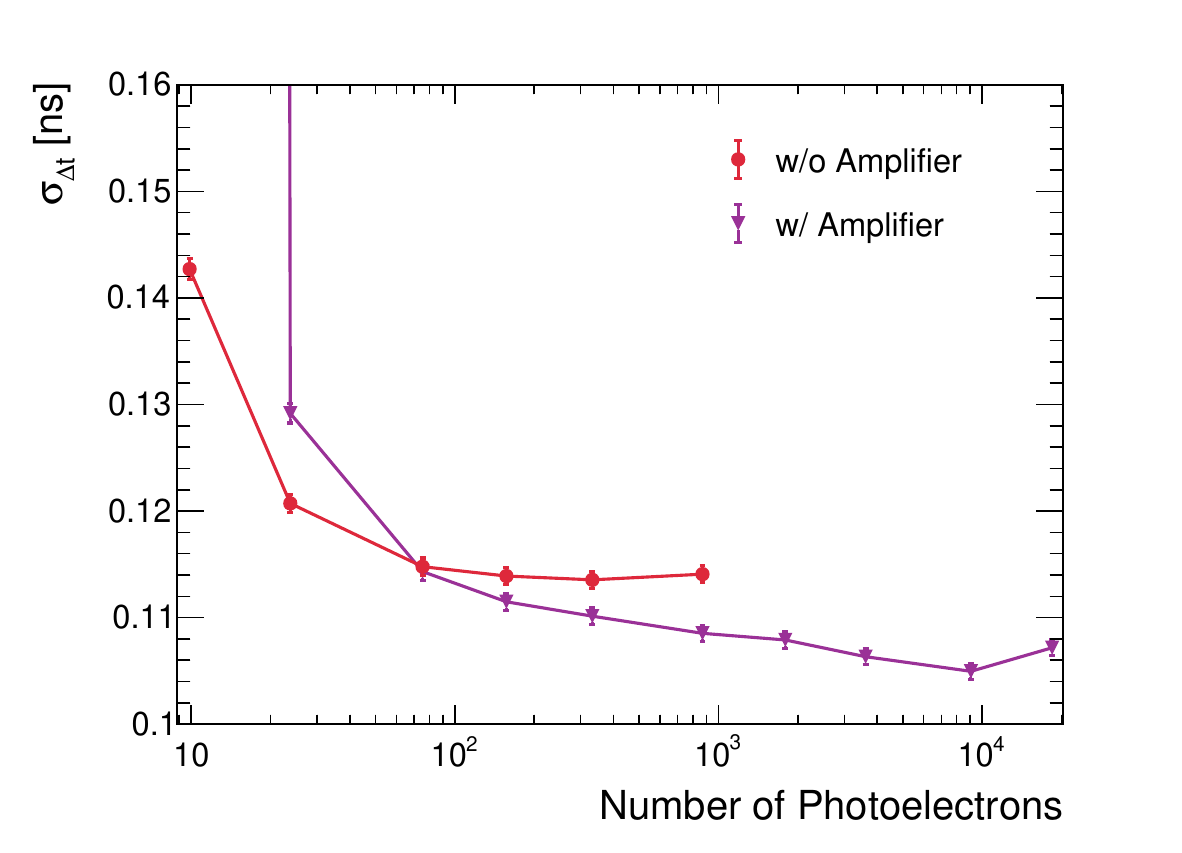}} 
    \caption{\label{fig:TR_electronics}~(a) Schematic of the electronic time resolution measurement. The laser pulse is received by the SiPM, and the amplified signal is sent to the oscilloscope. The standard deviation of the time difference between the SiPM signal and the synchronous signal defines the electronic time resolution. (b) Intrinsic time resolution of electronic components, which includes SiPM, preamplifier and oscilloscope.}
\end{figure}

The complete crystal detection unit consists of the crystal, SiPM, preamplifier, and oscilloscope, all of which contribute to the overall time resolution. To better characterize the intrinsic time resolution of the crystal, we first measured the time resolution of the electronic components, including the SiPM, preamplifier, and oscilloscope.

The experimental setup is illustrated in Figure~\ref{fig:TR_electronics}(a). A picosecond laser (PicoQuant Taiko PDL M1, IB-405-B) with a pulse width of less than 50 ps was used as the light source. The laser pulse was detected by the SiPM and either amplified by the preamplifier or directly sent to the oscilloscope. Simultaneously, a synchronous signal from the laser was fed directly to the oscilloscope. The standard deviation of the time difference between the SiPM-measured signal and the synchronous signal defines the intrinsic time resolution of the electronics. A standard constant fraction timing method was applied for analysis.

The results, shown in Figure~\ref{fig:TR_electronics}(b), indicate that the electronic time resolution improves as the number of detected photoelectrons increases but eventually saturates at approximately 110 ps. As light intensity increases, more photoelectrons are detected per unit time, reducing the impact of electronic noise and enhancing timing precision. However, due to the oscilloscope's 2.5 GS/s sampling rate, there is an upper limit to the achievable time resolution, leading to saturation. Additionally, a comparison of measurements with and without the preamplifier suggests that the intrinsic timing resolution contribution of the preamplifier is approximately 35 ps.

\section{Timing methods}

\begin{figure*}[h]
    \centering  
    \subfigure[]{
    \includegraphics[width=0.45\textwidth]{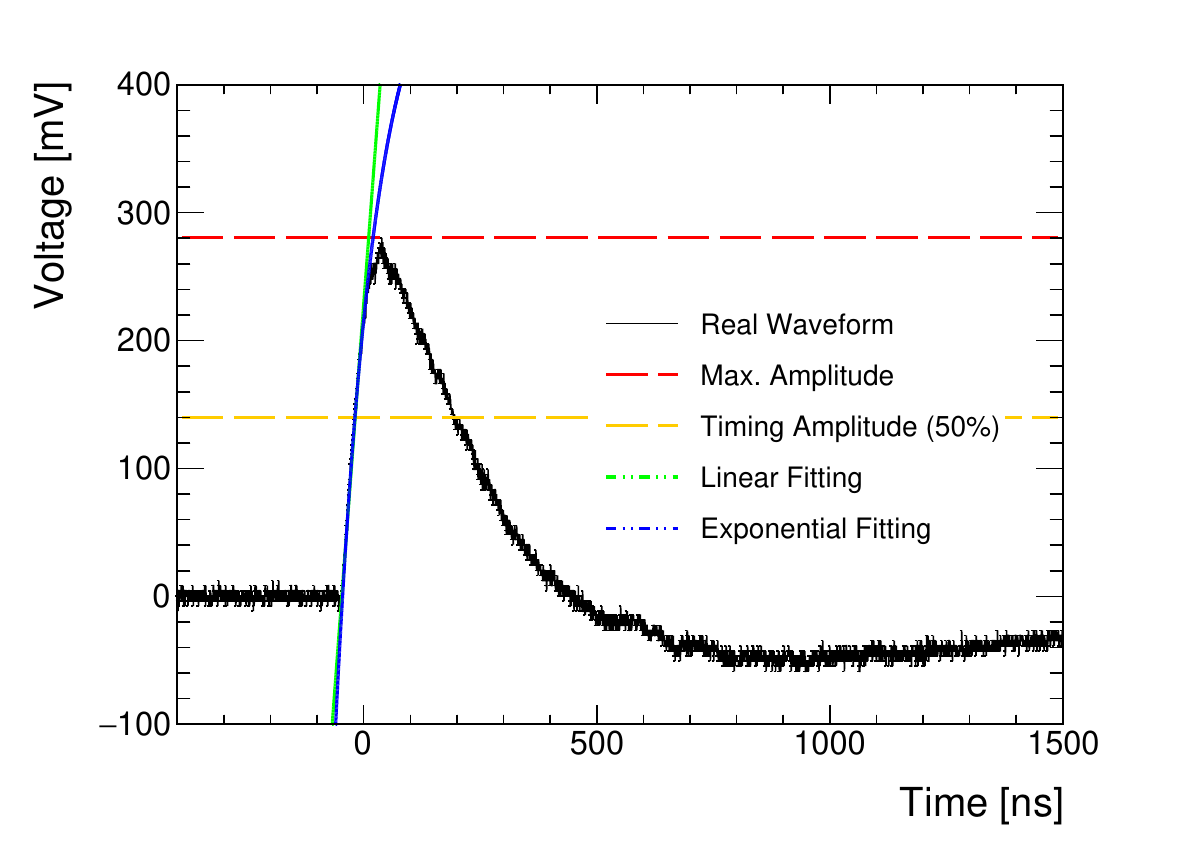}}
    \subfigure[]{
    \includegraphics[width=0.45\textwidth]{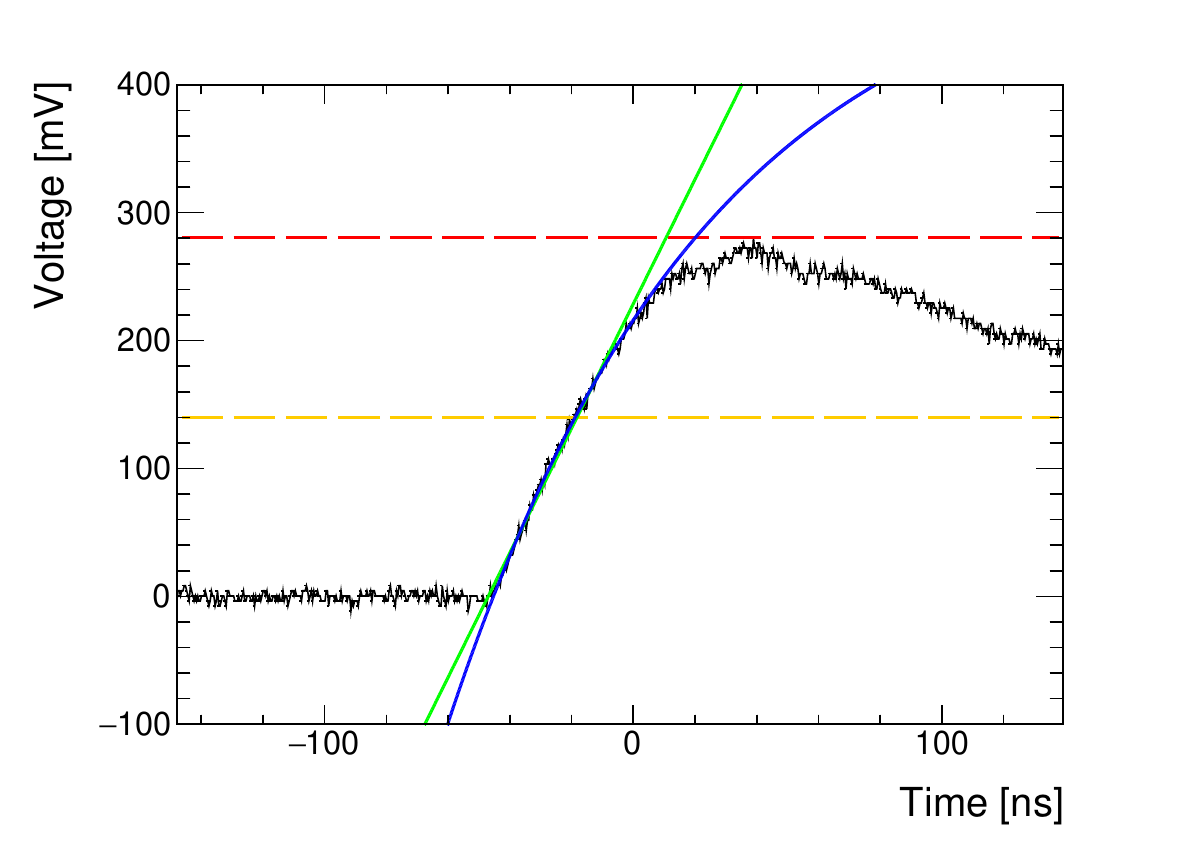}}
    \subfigure[]{
    \includegraphics[width=0.45\textwidth]{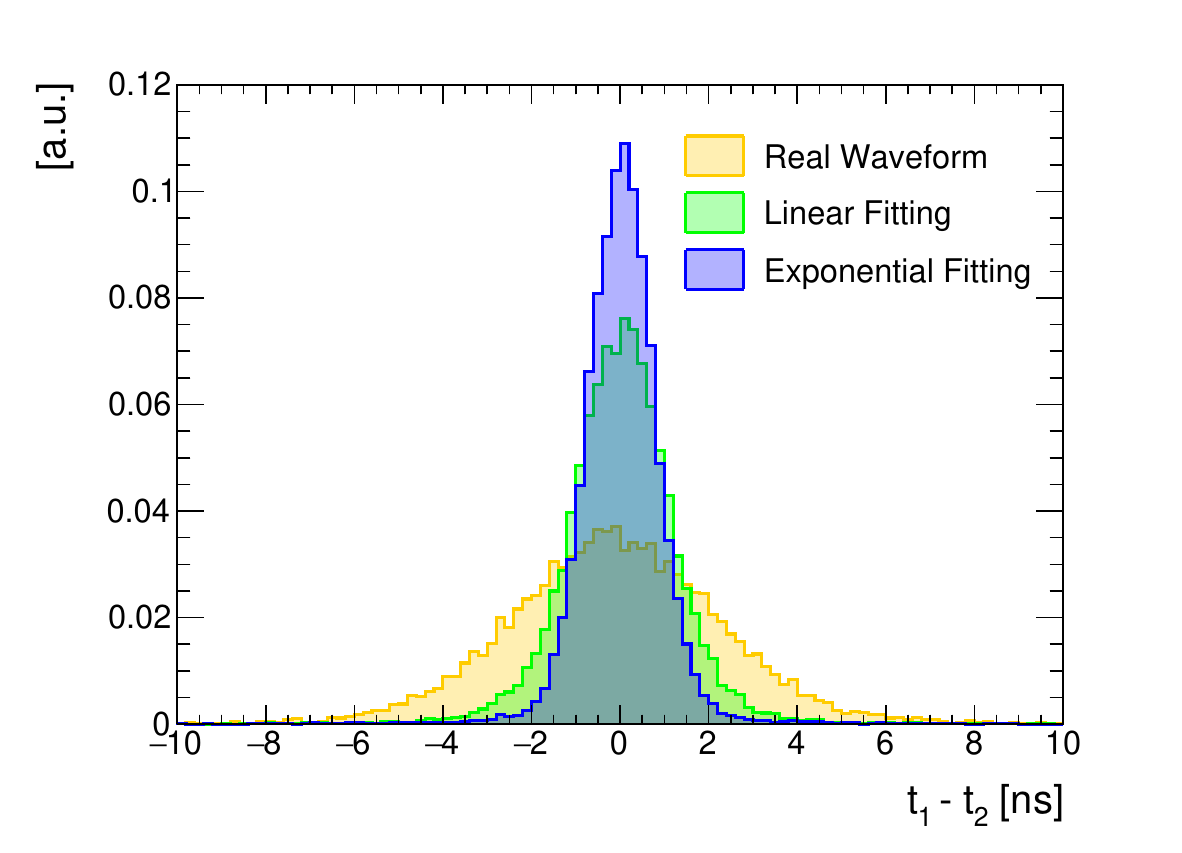}}
    \subfigure[]{
    \includegraphics[width=0.45\textwidth]{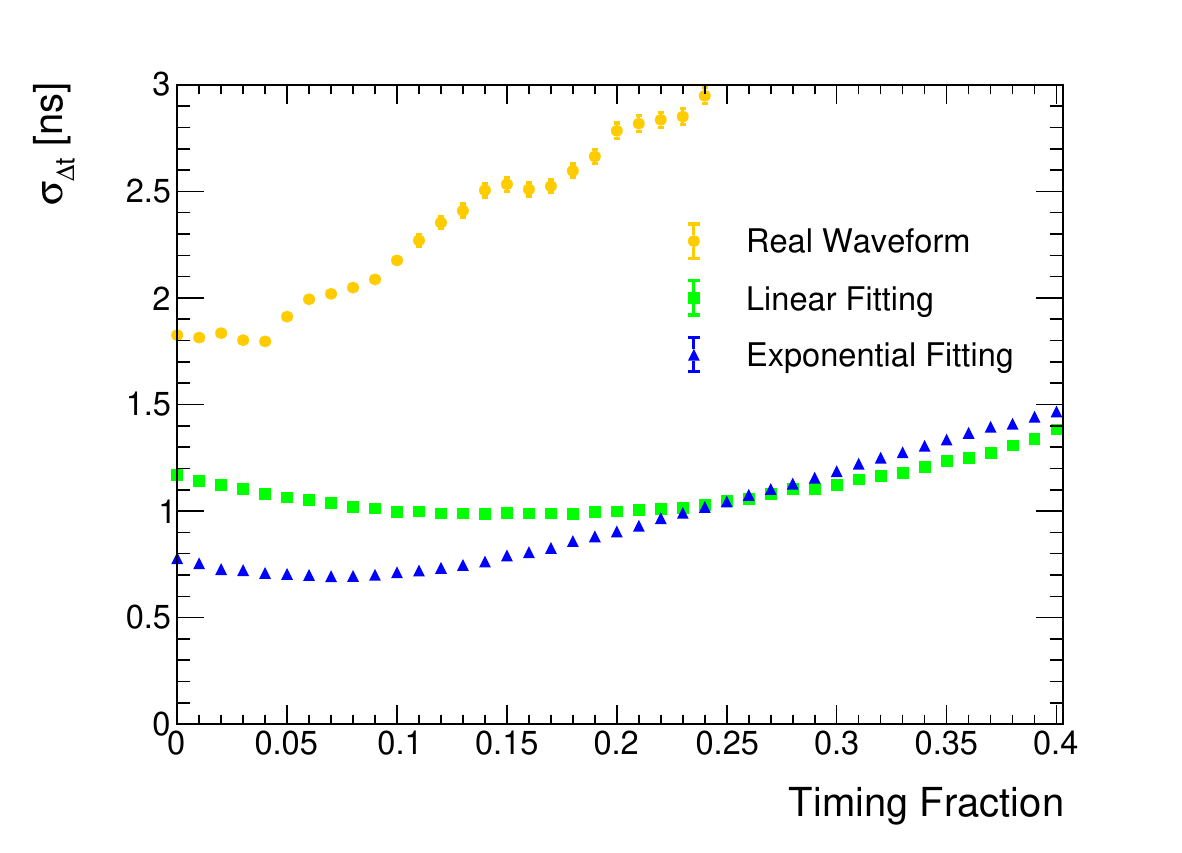}}
    \caption{\label{fig:TimingMethod}~(a) Signal waveform of scintillation light measured by the SiPM for a 1$\times$1$\times$40 cm$^3$ BGO crystal under 10 GeV $\pi^-$ beam, along with three different timing methods: constant fraction timing, linear fit of the rising edge, and exponential fit of the rising edge. (b) Detailed view of the waveform. The exponential function provides a better fit for the rising edge. (c) Time difference distributions between SiPM signals at both ends of the crystal bar for the three timing methods. A narrower distribution indicates better time resolution. (d) Time resolution as a function of the timing fraction. The exponential fit combined with around 10\% constant fraction timing yields the best timing performance.}
\end{figure*}

Figures~\ref{fig:TimingMethod}(a) and (b) present a typical signal waveform output from the SiPM positioned at one end of the crystal bar. The waveform exhibits a rapid rising edge followed by a slower falling edge, with an overshoot on the trailing edge due to the preamplifier. To determine the timing of the waveform, three methods were evaluated.

The first method is the classical constant fraction timing (CFT) approach. In this method, the maximum amplitude of the waveform is first identified. A point on the rising edge is then located at a fixed fraction of this maximum amplitude. The time at this point is defined as $t_1$, representing the timing for the signal from one end of the crystal. Similarly, $t_2$ is defined for the other end. The time resolution of the crystal-SiPM unit is determined as the standard deviation of the $t_1 - t_2$ distribution (blue histogram in Figure~\ref{fig:TimingMethod}(c), using a timing fraction of 10\%).

In addition to applying CFT directly to sampled points on the waveform, fitting functions can be employed to reduce the impact of noise on timing accuracy. Two fitting functions were tested for the steep region of the rising edge: a linear function and an exponential function. The CFT point was determined from the fitted function, and the resulting time difference distributions are shown in green and orange in Figure~\ref{fig:TimingMethod}(c) (both using a timing fraction of 10\%). The results indicate that the exponential fit provides the best time resolution at this timing fraction.

The choice of the timing fraction also influences timing precision. As shown in Figure~\ref{fig:TimingMethod}(d), the time resolution obtained from both linear and exponential fits initially improves with increasing timing fraction but then degrades beyond an optimal point. At lower timing fractions, the signal amplitude at the selected timing point is small, making the measurement more susceptible to electronic noise. As the timing fraction increases, the amplitude rises, reducing noise effects and improving time resolution. However, beyond a certain fraction, the slope of the waveform’s rising edge decreases, leading to greater statistical fluctuations and increased noise effects, which ultimately degrade time resolution.

Figure~\ref{fig:TimingMethod}(d) also shows that among the three timing methods, the exponential fit consistently provides the best time resolution. The optimal timing fraction is around 10\%, where the best timing performance is achieved. Based on these results, the exponential fit combined with a 10\% constant fraction timing method is used for subsequent analyses.

\section{Results}

\subsection{Energy response}

The time resolution of a crystal unit is primarily influenced by its light yield and scintillation decay time. Light yield quantifies the number of scintillation photons produced per unit energy deposited in the crystal, while scintillation decay time describes the rate at which the emitted light diminishes. A higher light yield and a shorter scintillation decay time can result in more detected photoelectrons per unit time, reducing timing jitter and improving time resolution. 

Under a 10 GeV pion beam, the effective MIP light yield and the rise time of the signal waveforms were measured for the crystals listed in Table~\ref{tab:Crystals}, each coupled with two SiPMs positioned at opposite ends of the crystal bar. The rise time of the waveform is defined as the time interval between the points on the rising edge corresponding to 10\% and 90\% of the maximum amplitude, which is primarily dictated by the scintillation decay characteristics of the crystal.

\begin{figure*}[h]
    \centering
    \includegraphics[width=0.7 \linewidth]{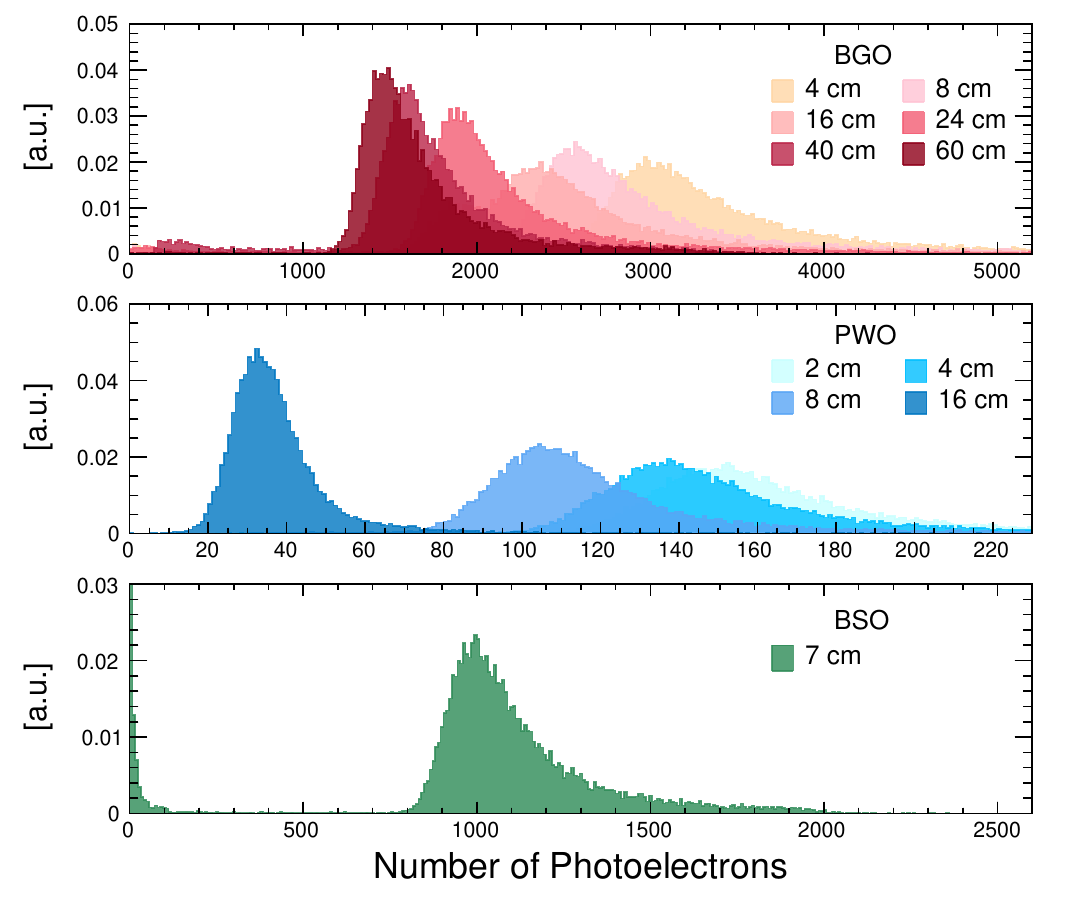}
    \caption{\label{fig:NPE_Crystals}~Distributions of photoelectrons detected by the SiPMs at both ends of the crystal bar under 10 GeV pion beam. From top to bottom, the red, green, and blue graphs correspond to the results for BGO, PWO, and BSO crystals, respectively. The legend values represent the crystal lengths, as listed in Table~\ref{tab:Crystals}.}
\end{figure*}

\begin{figure*}[h]
    \centering  
    \subfigure[]{
    \includegraphics[width=0.45\textwidth]{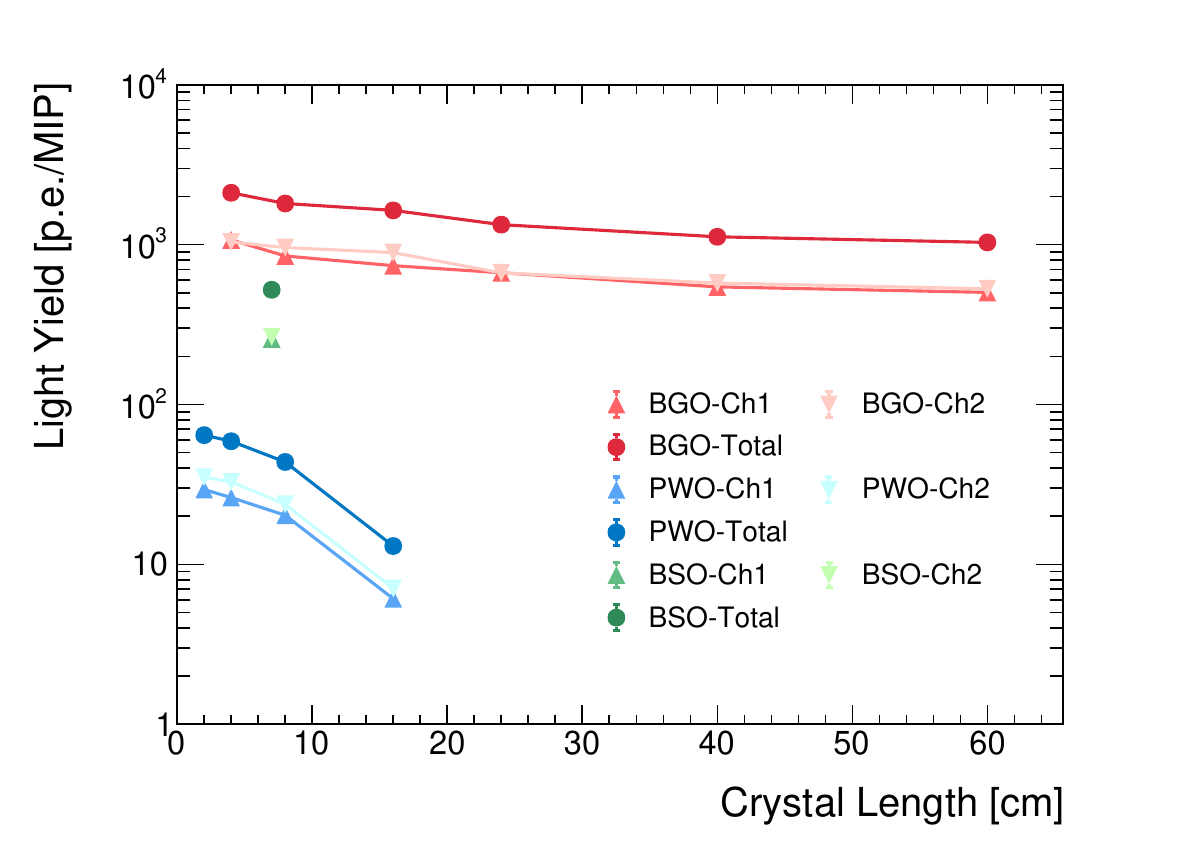}} 
    \subfigure[]{
    \includegraphics[width=0.45\textwidth]{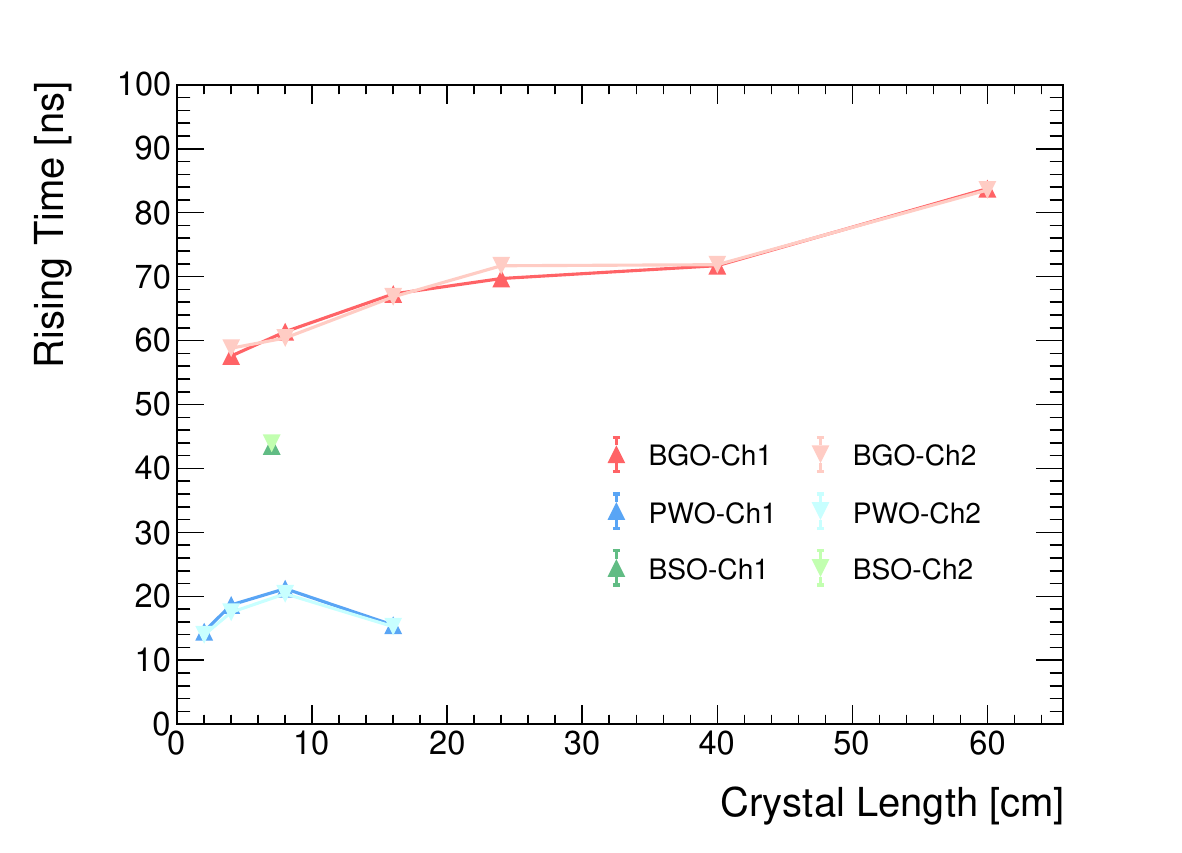}} 
    \caption{\label{fig:BGO_LY_tau}~(a) MIP light yield for different crystal units under 10 GeV pion beam. The horizontal axis represents the crystal lengths listed in Table~\ref{tab:Crystals}. Ch1, Ch2, and Total correspond to signals from the left-end SiPM, right-end SiPM, and the sum of both ends' SiPMs, respectively. (b) Rise time of the waveform’s rising edge for different crystal units under MIP signals. The rise time is defined as the time interval between the points on the rising edge corresponding to 10\% and 90\% of the maximum amplitude.}
\end{figure*}

Figure~\ref{fig:NPE_Crystals} presents the distributions of detected photoelectrons for different crystal units under 10 GeV pion beam. These distributions generally follow a Landau distribution, which characterizes the energy deposition by MIP particles, convolved with a Gaussian distribution representing the detector response. By fitting these distributions with a Landau-Gaussian convolution function, the most probable value (MPV) is extracted as the effective MIP light yield of each crystal unit.

Figure~\ref{fig:BGO_LY_tau}(a) compares the effective MIP light yields for BGO, PWO, and BSO crystal units. Among the three types of crystals, BGO exhibits the highest light yield. For crystals of similar dimensions, the light yield of BGO is approximately four times that of BSO and 40 times that of PWO. Within a given crystal type, the light yield decreases as crystal length increases due to self-absorption effects. Longer crystals require scintillation light to propagate over greater distances, leading to increased attenuation and reduced light collection at the SiPMs.

Figure~\ref{fig:BGO_LY_tau}(b) shows the rise times of signal waveforms for different crystal units. BGO, with the slowest scintillation decay time, exhibits the longest rise time, exceeding 50 ns. PWO, in contrast, has the shortest rise time, below 20 ns, while BSO falls in between. Within the same crystal type, the rise time generally increases with crystal length due to longer light propagation distances. An exception is observed for the 16 cm PWO crystal, where low light yield results in an insufficient number of detected photoelectrons, limiting the ability to accurately reflect the crystal’s scintillation properties.

\subsection{Time resolution measured with pions}

As shown in the photoelectron spectra in Figure~\ref{fig:NPE_Crystals}, the energy deposition of a 10 GeV pion in the crystal exhibits a long tail. To study the time resolution of the crystal unit under a 1-MIP signal, events with photoelectron counts between 0.5 and 1.5 times the MIP light yield were selected.

\subsubsection{Time resolution of different crystals}

\begin{figure}[h]
    \centering
    \includegraphics[width=0.9\linewidth]{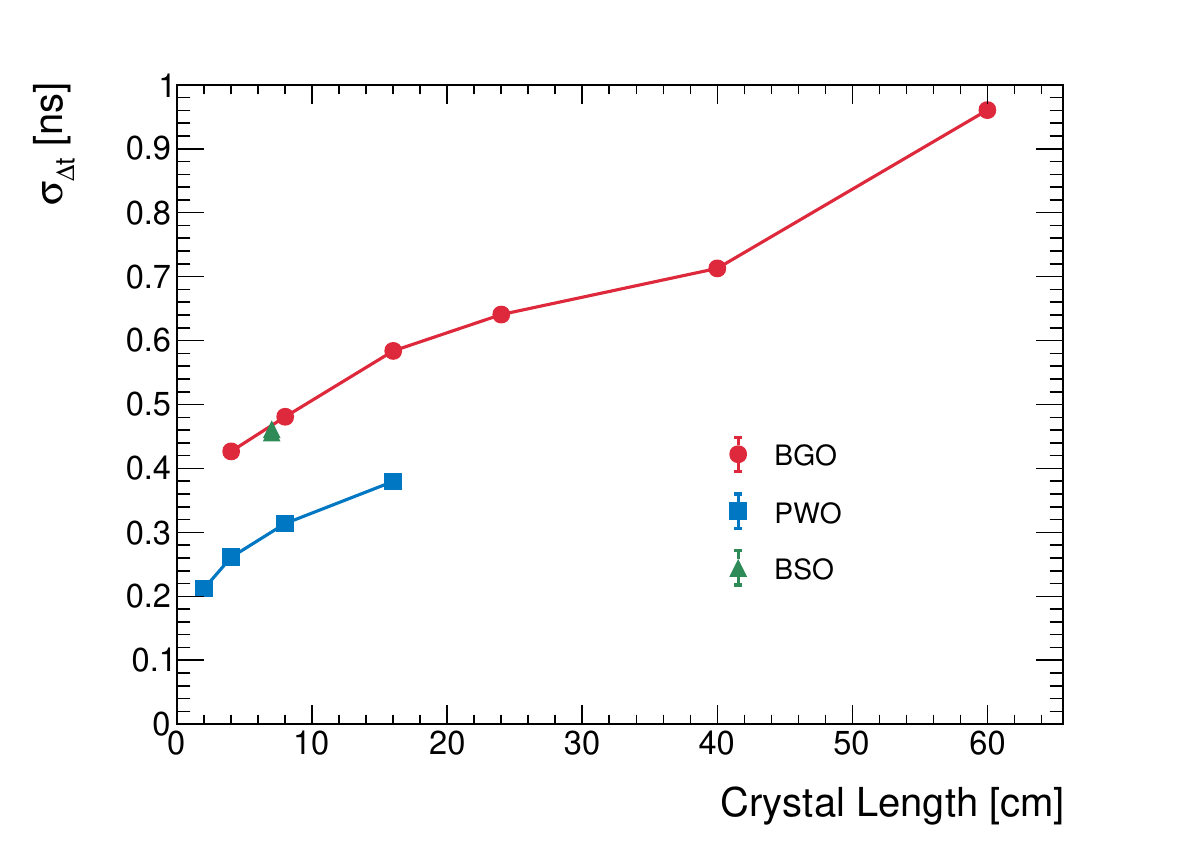}
    \caption{\label{fig:TR_MIP}~MIP time resolution for BGO, BSO, and PWO crystal units of different sizes under 10 GeV pion beam. The selected events include cases where the total photoelectrons detected by the tw0 SiPMs at both ends of the crystal bar range from 0.5 to 1.5 times the MIP light yield.}
\end{figure}

The results in Figure~\ref{fig:TR_MIP} were obtained with particles controlled to hit the center of the crystal bar. For each type of crystal, the MIP time resolution degrades as the crystal length increases. This deterioration is primarily attributed to reduced light yield, slower signal rise times, and fewer detected photoelectrons per unit time in longer crystals.

The time resolution of the crystal unit is influenced by both the light yield and the scintillation decay time, which is reflected in the signal rise time among the crystals in Figure~\ref{fig:BGO_LY_tau}. For crystals with lengths of 7–8 cm, PWO exhibits the best time resolution. Despite its low light yield, its extremely fast scintillation decay time enables a rapid signal rise, leading to superior timing performance. In contrast, BGO and BSO display opposite characteristics in terms of light yield and scintillation decay time. As a result, the time resolutions of BGO and BSO crystals of similar dimensions are nearly identical.

\subsubsection{Time resolution of long crystal bars}

For long crystal bars, the signals detected by the SiPMs at both ends are attenuated, and the signal amplitude may vary depending on the particle's hit position along the crystal. This can lead to non-uniform energy and time responses, degrading the overall time resolution of the crystal unit. Additionally, factors such as crystal growth, machining, packaging, and temperature variations can further contribute to response inconsistencies.

\begin{figure}[h]
    \centering  
    \subfigure[]{
    \includegraphics[width=0.45\textwidth]{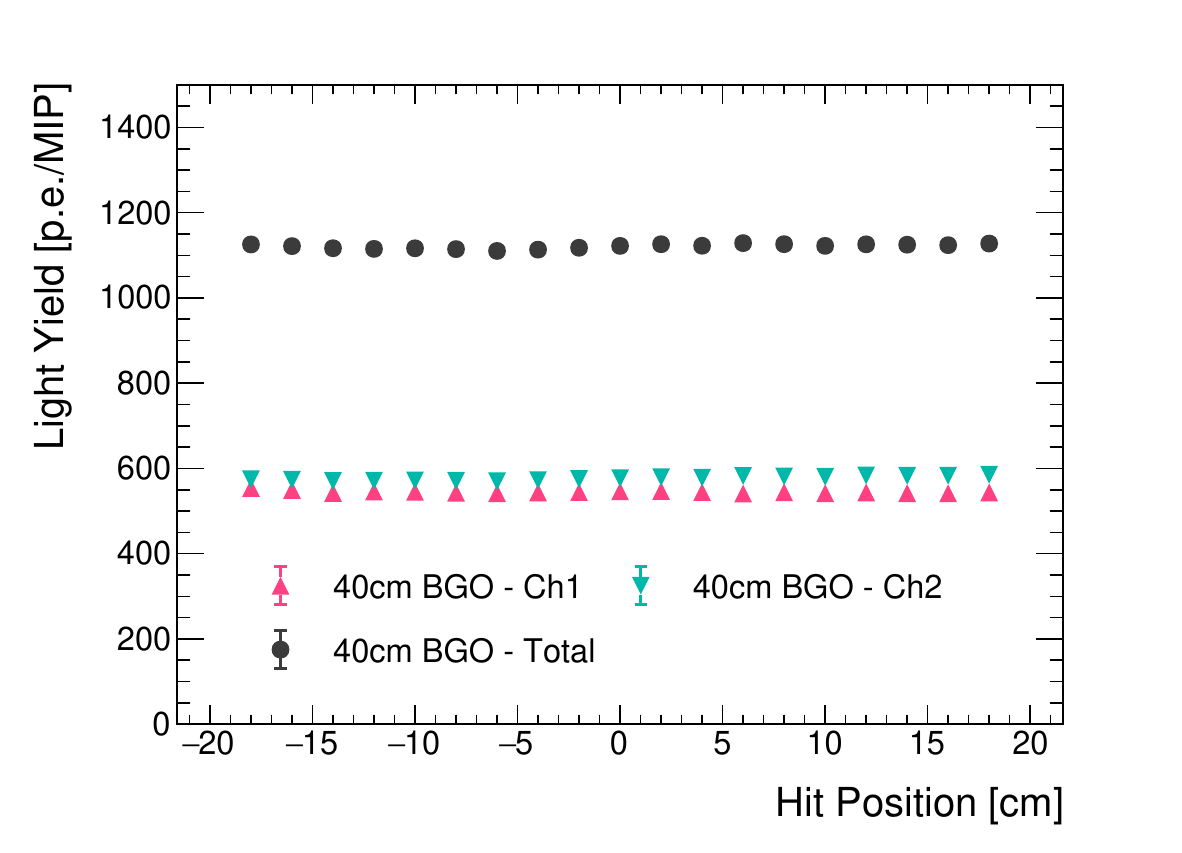}}
    \subfigure[]{
    \includegraphics[width=0.45\textwidth]{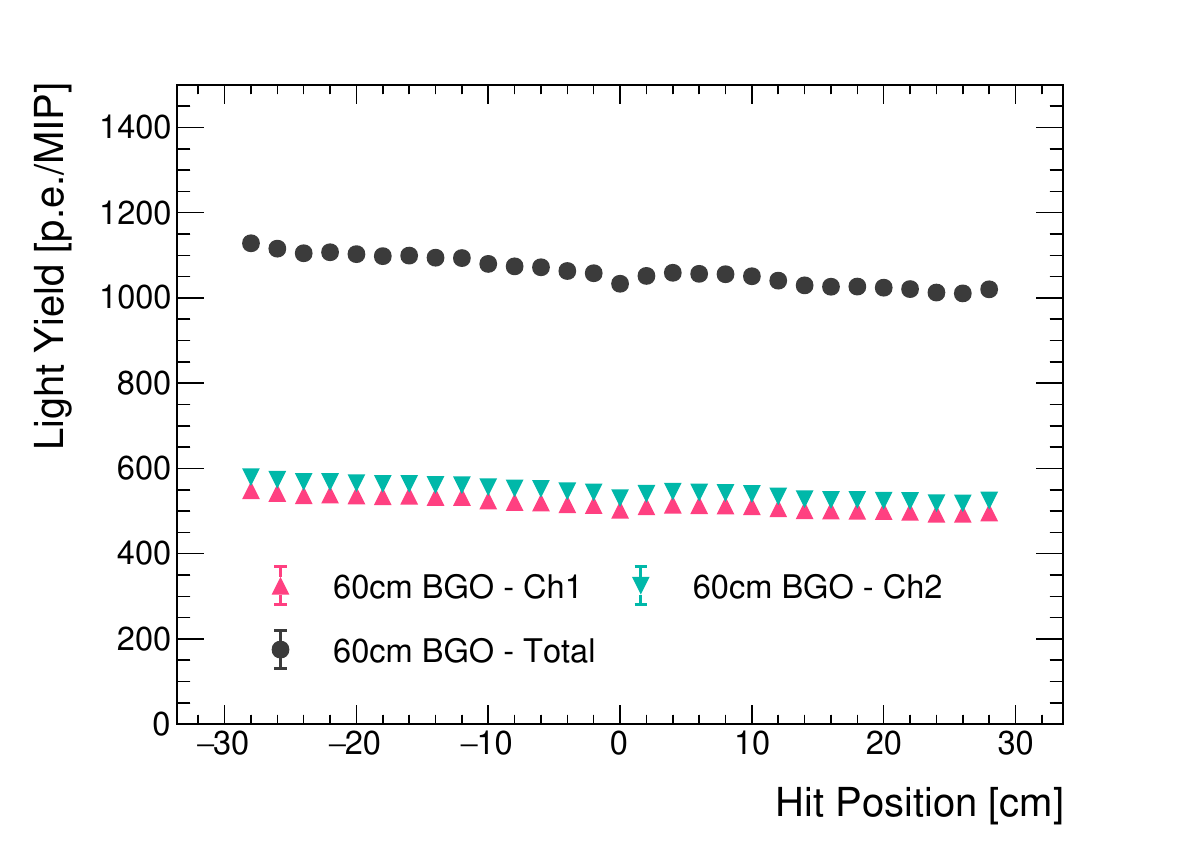}}
    \caption{\label{fig:MIP_Scan_LY}~Light yield as a function of particle hit position for (a) 1$\times$1$\times$40 cm$^3$ and (b) 1.5$\times$1.5$\times$60 cm$^3$ BGO crystal units. The horizontal axis represents the particle hit position along the crystal length. The incident particles are 10 GeV pions.}
\end{figure}

To evaluate the uniformity of energy and time responses, pion beams were used to scan two long BGO crystal bars with dimensions of 1$\times$1$\times$40 cm$^3$ and 1.5$\times$1.5$\times$60 cm$^3$ along their lengths. The MIP light yields measured at different hit positions are shown in Figure~\ref{fig:MIP_Scan_LY}. For the 40 cm BGO crystal unit, both the single-end signals and the sum of signals from both ends exhibited excellent uniformity, with variations within 1\% (Table~\ref{tab:LY_Uniformity}). Here, the uniformity of MIP light yield is defined as the standard deviation of the light yield across all hit positions divided by the average value. In contrast, the 60 cm BGO crystal unit displayed worse uniformity, within 3\%, showing a gradual decrease from one end to the other, likely due to non-uniformities introduced during crystal growth and machining.

\begin{table}[width=.9\linewidth,h]
\centering
\fontsize{8}{12}\selectfont
\caption{\label{tab:LY_Uniformity}~MIP light yield uniformity along the length of BGO crystal bars. Uniformity is defined as the standard deviation of the crystal unit's response to particles at different hit positions, divided by the average value.}
    \begin{tabular}{cccc}
        \toprule
        \makecell[c]{Crystal dimension} &\makecell[c]{Channel-1} &\makecell[c]{Channel-2} &\makecell[c]{Total}\\ 
        \midrule
        1$\times$1$\times$40 cm$^3$ & 0.44\% & 0.88\% & 0.43\% \\
        1.5$\times$1.5$\times$60 cm$^3$ & 2.9\% & 3.0\% & 3.0\% \\
        \bottomrule
    \end{tabular}
\end{table}

\begin{figure}[h]
    \centering
    \includegraphics[width=0.9\linewidth]{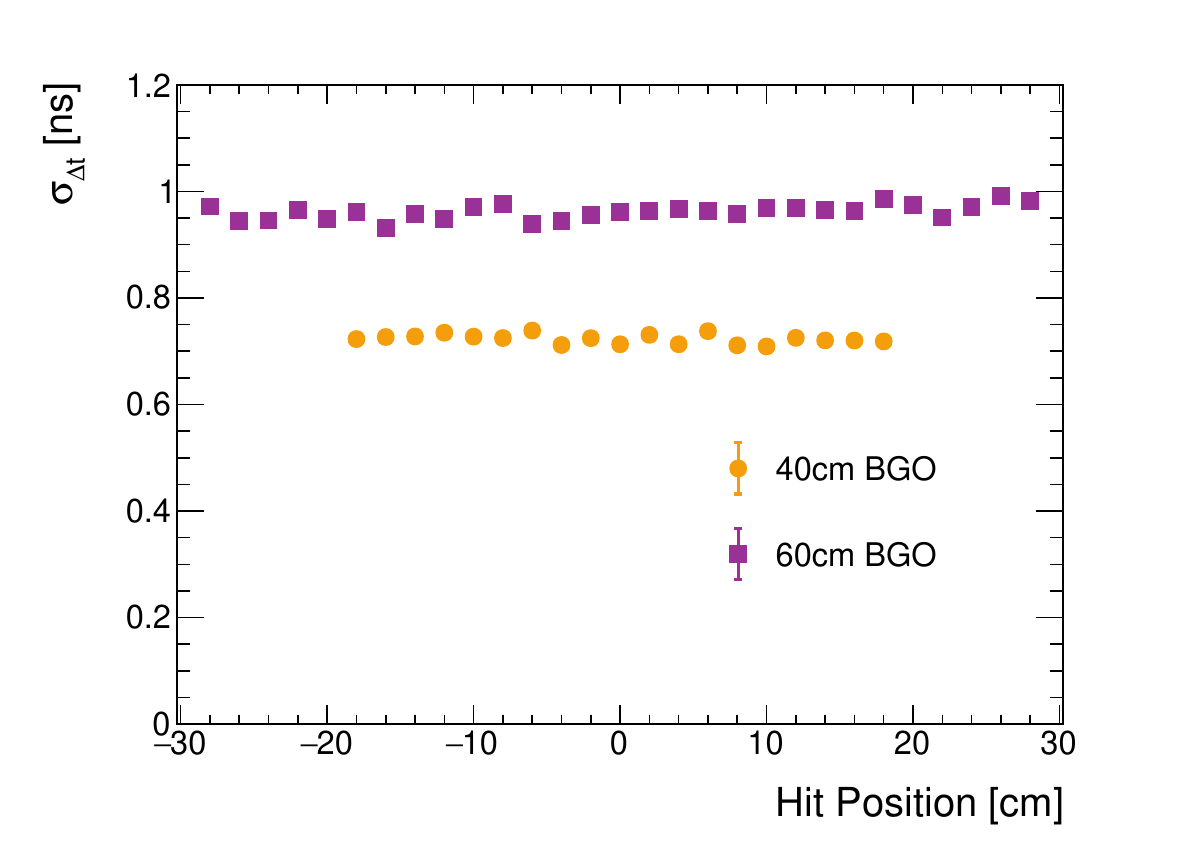}
    \caption{\label{fig:MIP_Scan_Time}~Time resolution as a function of particle hit position for the 1$\times$1$\times$40 cm$^3$ and 1.5$\times$1.5$\times$60 cm$^3$ BGO crystal units. The incident particles are 10 GeV pions, hitting different positions along the crystal length.}
\end{figure}

The relationship between time resolution and particle hit position is shown in Figure~\ref{fig:MIP_Scan_Time}. Both the 40 cm and 60 cm BGO crystals demonstrated good time resolution uniformity. The non-uniformity in light yield for the 60 cm BGO crystal shown in Figure~\ref{fig:MIP_Scan_LY}(b) did not significantly impact its timing performance. Overall, the time resolution for the 40 cm BGO crystal unit was approximately 0.75 ns for a 1-MIP signal, while the 60 cm BGO crystal unit achieved a time resolution of about 0.95 ns.

\begin{figure}[htbp]
    \centering  
    \subfigure[]{
    \includegraphics[width=0.45\textwidth]{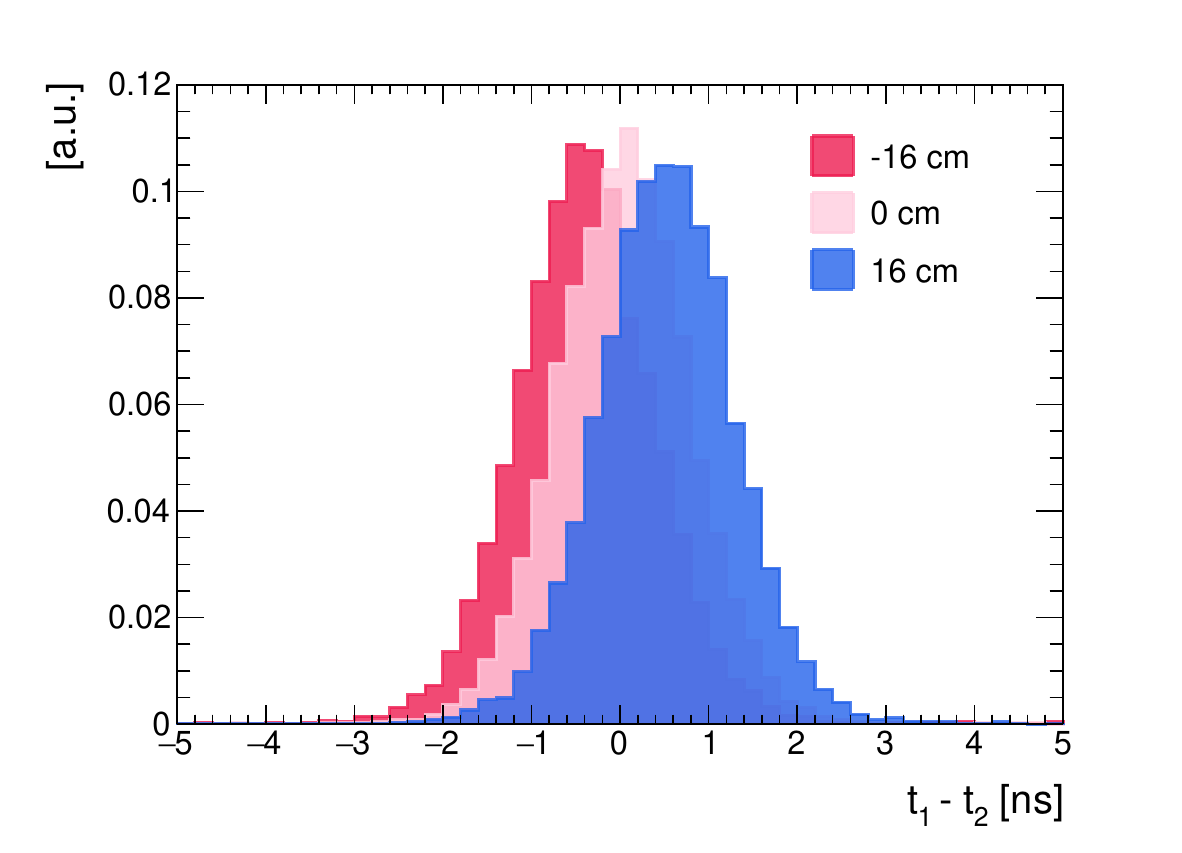}}
    \subfigure[]{
    \includegraphics[width=0.45\textwidth]{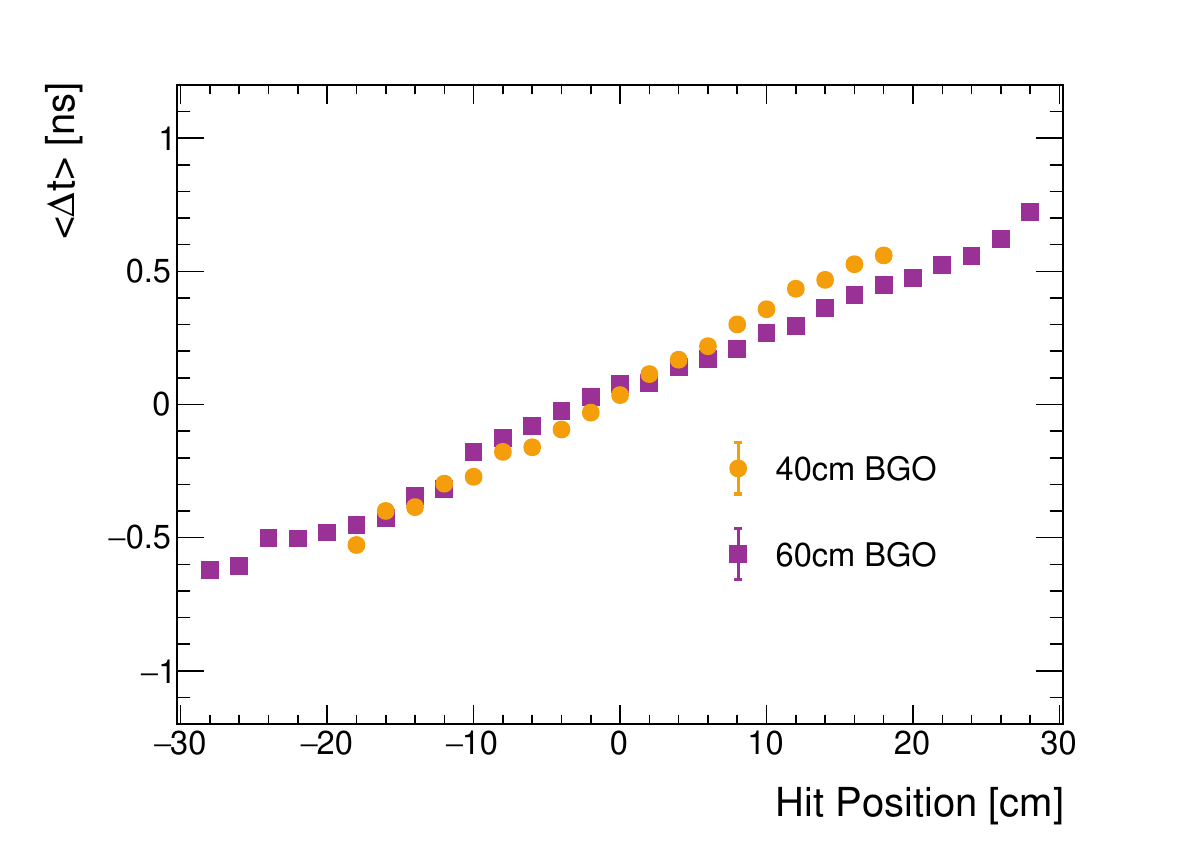}}
    \caption{\label{fig:MIP_Scan_TimeDiff}~(a) Time difference distribution of SiPM readout signals at both ends of the 1$\times$1$\times$40 cm$^3$ BGO crystal bar. The red, pink, and blue histograms represent the distributions for 10 GeV pions incident at positions -16 cm, 0 cm, and 16 cm along the crystal length. (b) Time difference as a function of particle hit position for the 1$\times$1$\times$40 cm$^3$ and 1.5$\times$1.5$\times$60 cm$^3$ BGO crystal units.}
\end{figure}

The time difference between the signals from the two SiPMs can be used to determine the particle hit position along the crystal bar. Figure~\ref{fig:MIP_Scan_TimeDiff}(a) shows the time difference distributions for three hit positions relative to the center of the 40 cm BGO crystal unit: -16 cm, 0 cm, and +16 cm. The SiPM closer to the hit position responds earlier, causing the mean value of the time difference distribution to vary systematically with the hit position. As shown in Figure~\ref{fig:MIP_Scan_TimeDiff}(b), this variation follows an approximately linear trend. 

Based on the measured time resolution results in Figure~\ref{fig:MIP_Scan_LY}, the position resolution for the 40 cm BGO crystal unit using timing information from both ends is estimated to be approximately 25 cm, while the position resolution for the 60 cm BGO crystal unit is about 40 cm.

\subsection{Time resolution measured with electrons}

\subsubsection{Time resolution at different amplitudes}

In addition to evaluating the time resolution of crystal units under MIP signals, it is also crucial to assess their performance under high-energy signals. Unlike the minimum ionization process of pions, electrons induce EM showers, generating numerous secondary particles and depositing significantly more energy in the crystal. The depth and width of the EM shower depend on the material. For a given material, the energy deposition per unit thickness initially increases as the shower develops, peaks at the shower maximum, and then gradually decreases.

To control the energy absorbed by the test crystals, additional BGO crystals were placed upstream as a pre-shower, as shown in Figure~\ref{fig:BeamSetup}. By varying the pre-shower thickness, the test crystals were positioned at different shower depths. The incident electron beam was perpendicular to the crystal and passed through its center.

\begin{figure*}[h]
    \centering  
    \subfigure[]{
    \includegraphics[width=0.45\textwidth]{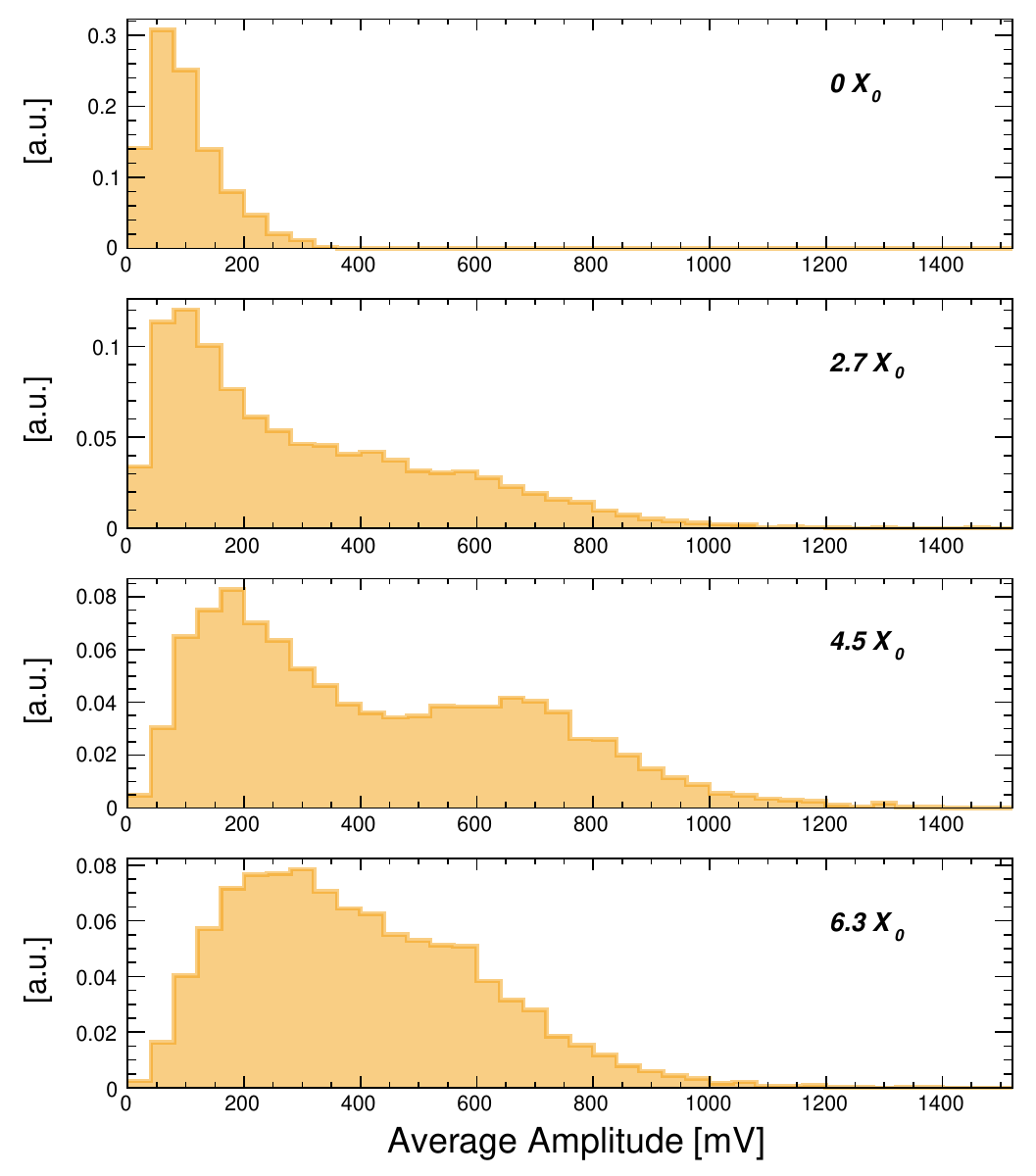}} 
    \subfigure[]{
    \includegraphics[width=0.45\textwidth]{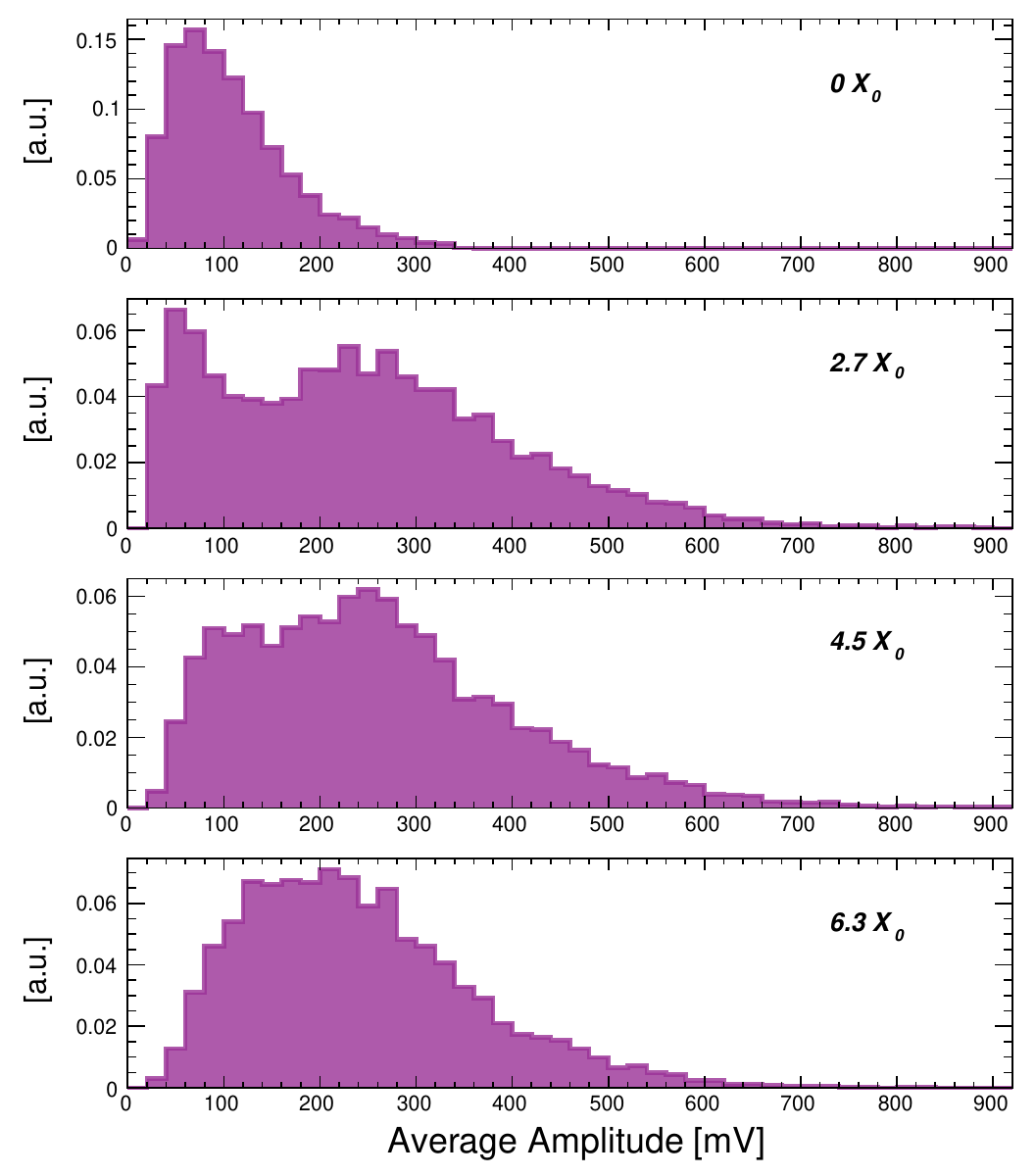}}
    \caption{\label{fig:Amp_pre-shower}~Average signal amplitude distribution of the electromagnetic shower from 5 GeV electrons measured by the (a) 1$\times$1$\times$40 cm$^3$ and (b) 1.5$\times$1.5$\times$60 cm$^3$ BGO crystal units under different pre-shower thicknesses. The average signal amplitude is defined as the mean of the amplitudes read out by the two SiPMs at both ends of the crystal.}
\end{figure*}

Figure~\ref{fig:Amp_pre-shower} shows the signal amplitude distributions for the 1$\times$1$\times$40 cm$^3$ and 1.5$\times$1.5$\times$60 cm$^3$ BGO crystal units. As the pre-shower thickness increases, the recorded signals grow larger, reaching a maximum at approximately 4.5 radiation lengths, after which the amplitude starts to decrease. However, in both Figures~\ref{fig:Amp_pre-shower}(a) and (b), a significant fraction of small signals around 100 mV is observed. This is due to the limited 2 cm thickness of the pre-shower crystals, which allows some beam particles to bypass the pre-shower and directly hit the test crystals.

\begin{figure*}[h]
    \centering  
    \subfigure[]{
    \includegraphics[width=0.45\textwidth]{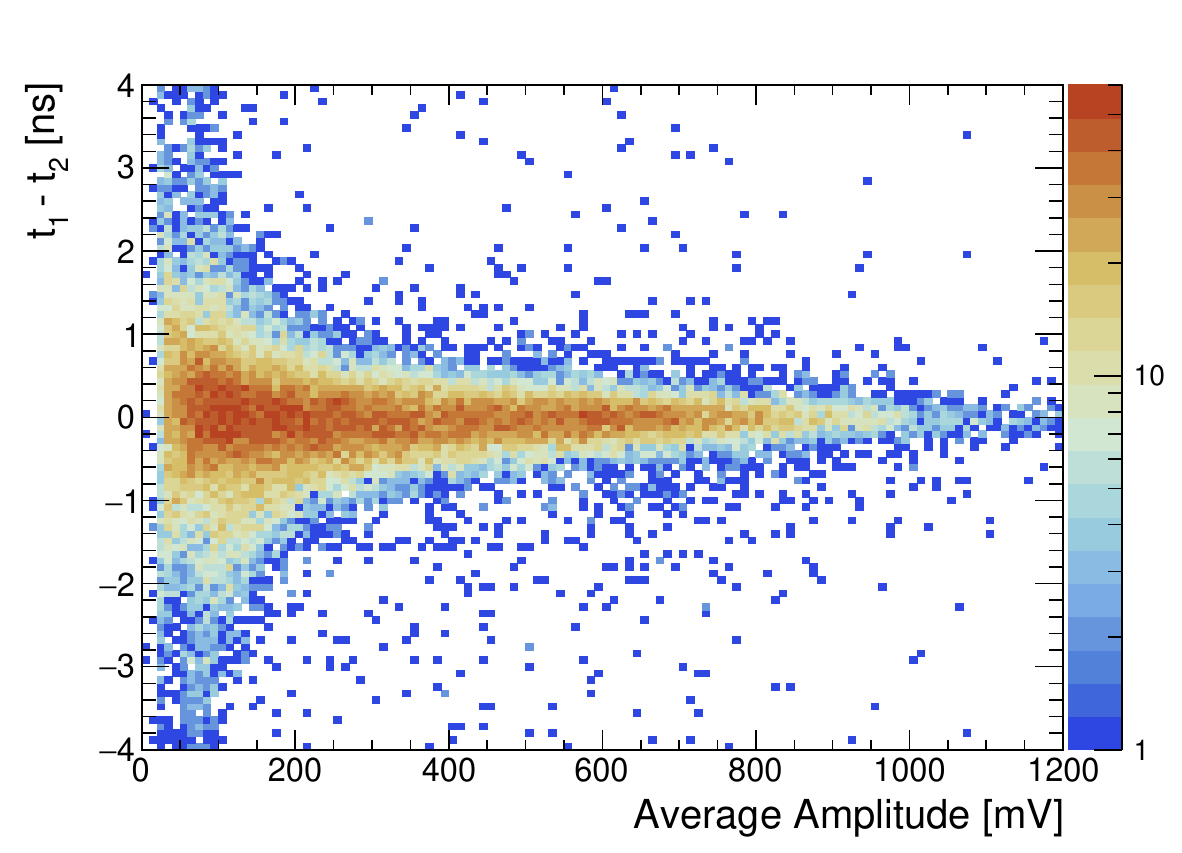}} 
    \subfigure[]{
    \includegraphics[width=0.45\textwidth]{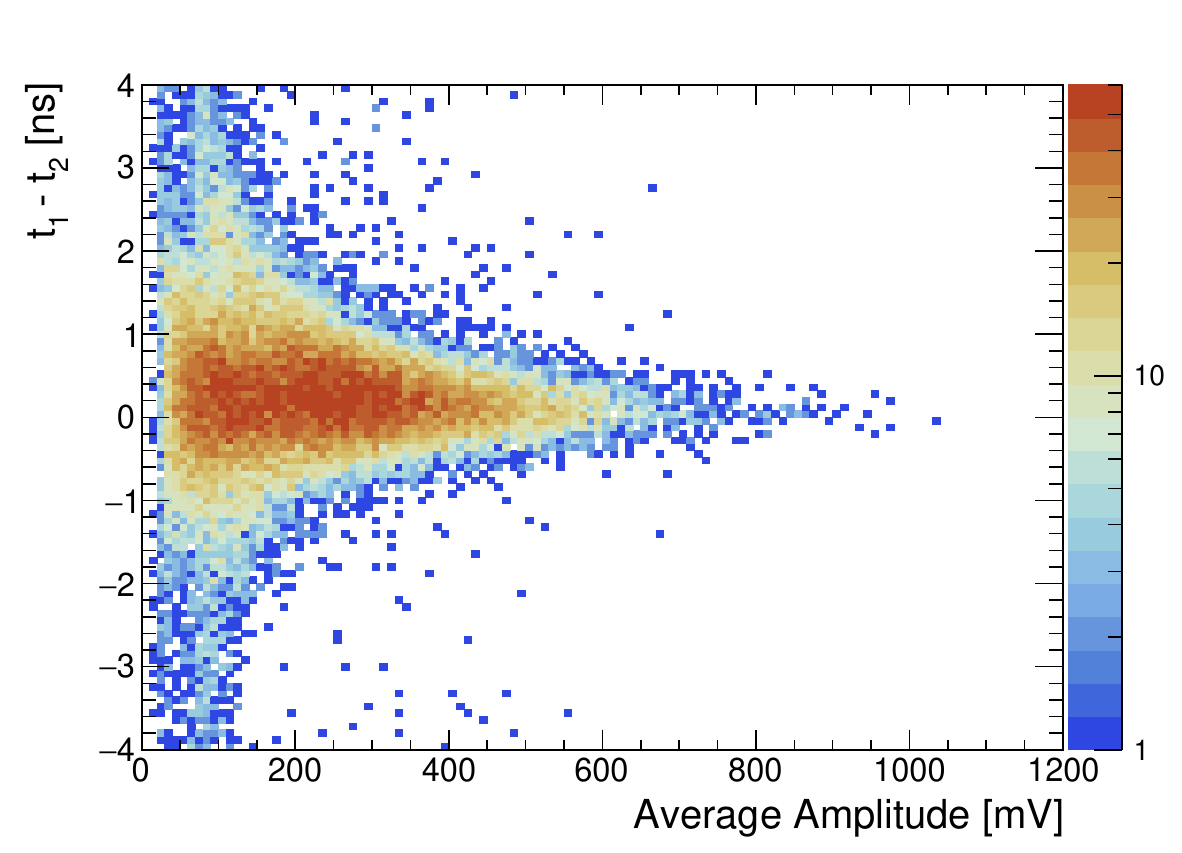}}
    \caption{\label{fig:Diff_pre-shower}~Two-dimensional distribution of time difference versus average signal amplitude measured by the (a) 1$\times$1$\times$40 cm$^3$ and (b) 1.5$\times$1.5$\times$60 cm$^3$ BGO crystal units. The data samples are combined from different pre-shower thicknesses shown in Figure~\ref{fig:Amp_pre-shower}.}
\end{figure*}

By combining the data from both crystal units in Figure~\ref{fig:Amp_pre-shower}, the relationship between the time difference of signals from the two ends of the crystal bar and the average signal amplitude is shown in Figure~\ref{fig:Diff_pre-shower}. The narrower time difference distributions for larger average signal amplitudes indicate improved timing precision for high-energy signals. This is because, in high-energy signals, the SiPM detects more photoelectrons per unit time, reducing statistical fluctuations in the waveform and minimizing time jitter. 

\begin{figure}[h]
    \centering
    \includegraphics[width=0.9\linewidth]{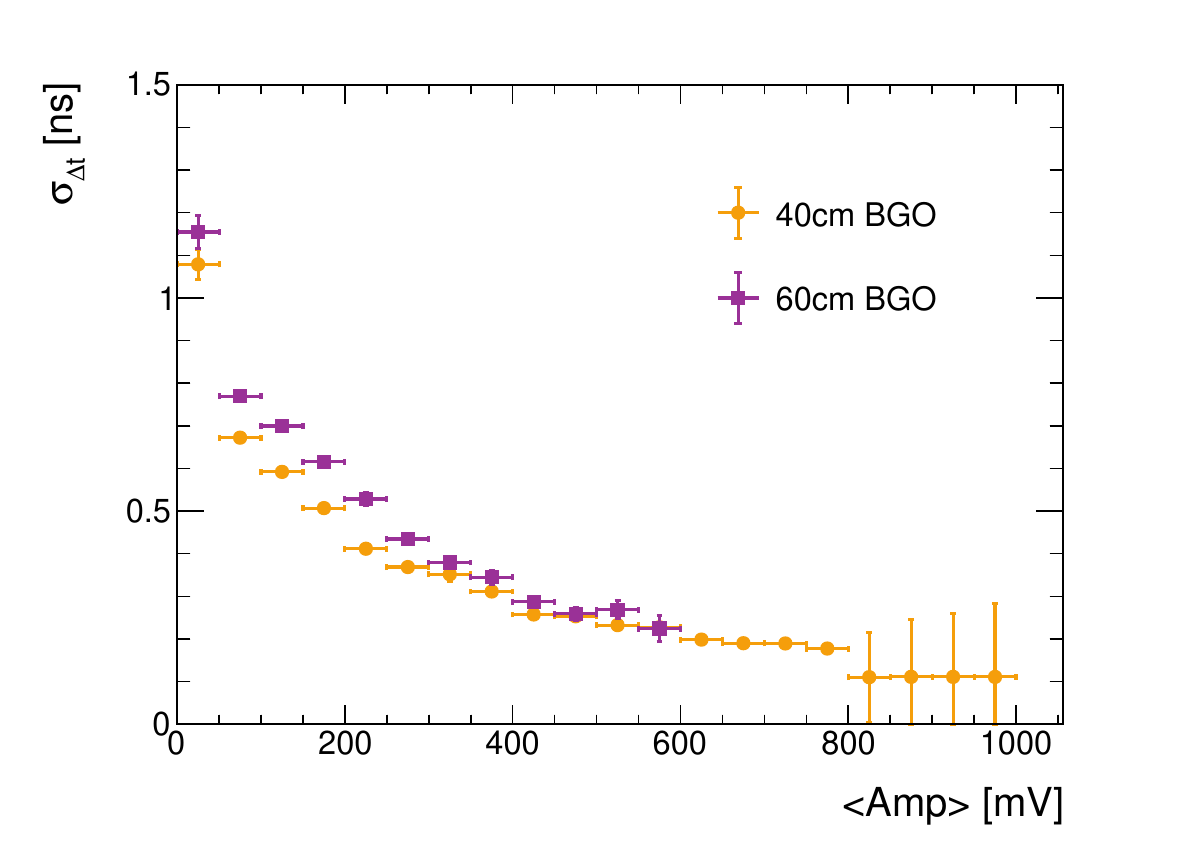}
    \caption{\label{fig:TR_electron}~Time resolution of the 1$\times$1$\times$40 cm$^3$ and 1.5$\times$1.5$\times$60 cm$^3$ BGO crystal units under electromagnetic showers induced by 5 GeV electrons, as a function of the average signal amplitude. Data from different pre-shower thicknesses are combined.}
\end{figure}

The dependence of time resolution on average signal amplitude is shown in Figure~\ref{fig:TR_electron}. For both BGO crystal units, the time resolution improves significantly with increasing amplitude. At extremely high signal amplitudes, the 60 cm BGO crystal unit achieves a time resolution of 220 ps, while the 40 cm unit performs slightly better under similar conditions. At an average amplitude of approximately 775 mV, equivalent to around 20 MIPs, the time resolution approaches 200 ps. For amplitudes exceeding 800 mV, the time resolution saturates due to limitations imposed by the oscilloscope's 2.5 GS/s sampling rate.

\subsubsection{Time resolution at different shower depth}

\begin{figure}[h]
    \centering  
    \subfigure[]{
    \includegraphics[width=0.45\textwidth]{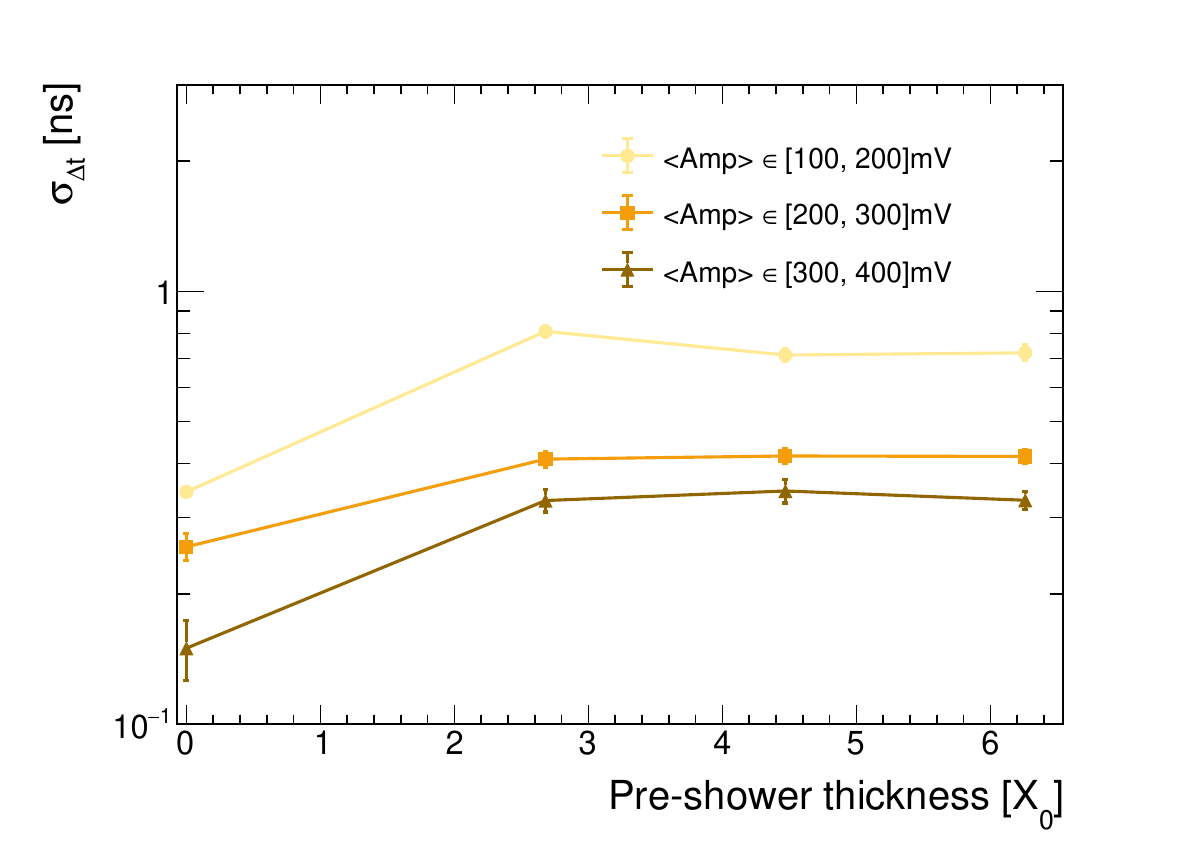}} 
    \subfigure[]{
    \includegraphics[width=0.45\textwidth]{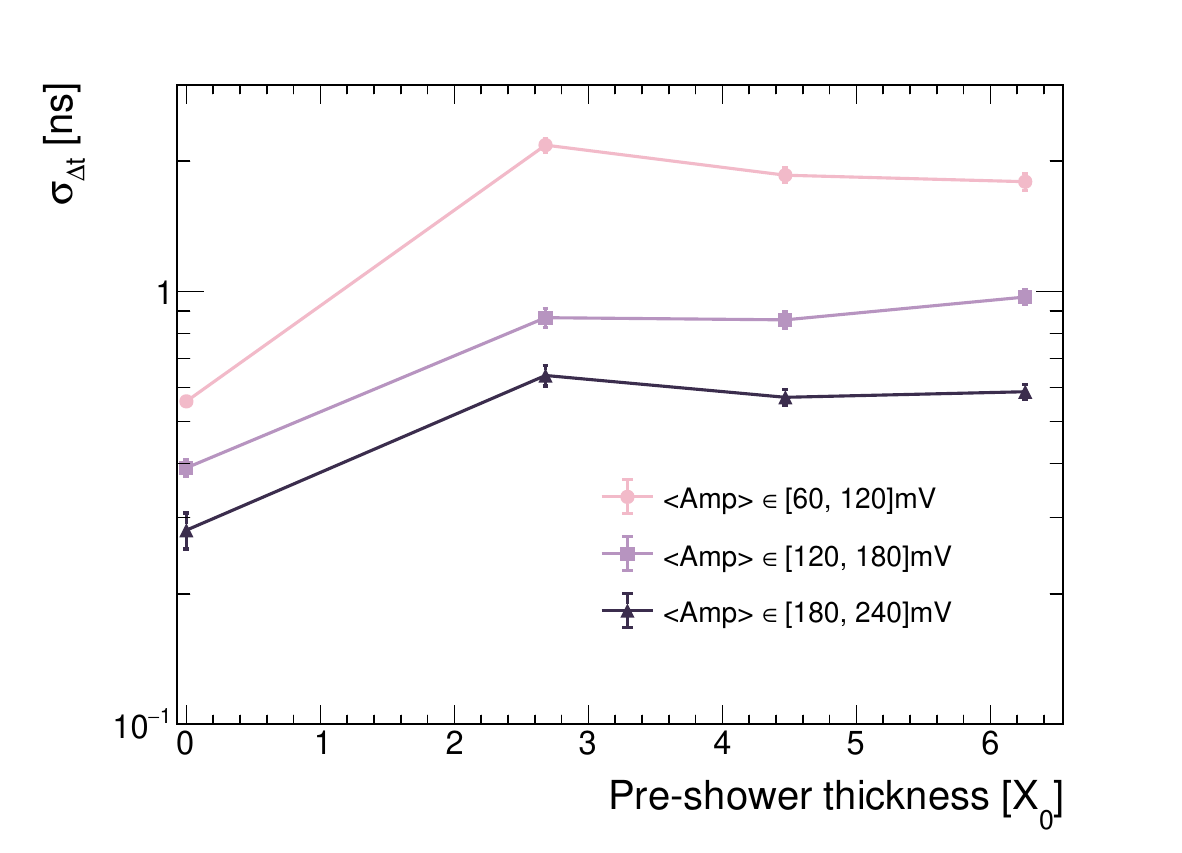}}
    \caption{\label{fig:TR_showerdepth}~Time resolution of the (a) 1$\times$1$\times$40 cm$^3$ and (b) 1.5$\times$1.5$\times$60 cm$^3$ BGO crystal units as a function of pre-shower thickness. Each set of data points corresponds to cases where the average signal amplitude is constrained within a 100 mV range to ensure comparable energy absorption across measurements.}
\end{figure}

During different stages of shower development, not only does the energy deposition vary, but the time resolution for the same energy deposition may also be affected. By selecting signals within a fixed amplitude range from data obtained under different pre-shower thicknesses, we can compare the impact of shower depth on time resolution. The results, shown in Figure~\ref{fig:TR_showerdepth}, exhibit similar trends for both the 40 cm and 60 cm BGO crystal units. For a fixed pre-shower thickness, the time resolution within the crystal unit improves with increasing signal amplitude, as higher-energy signals exhibit reduced time jitter at the rising edge, resulting in better timing accuracy.

For signals with similar amplitudes (below 100 mV) but different pre-shower thicknesses, a noticeable difference in time resolution is observed only between cases with and without a pre-shower. When a pre-shower of 2.7 radiation lengths is introduced, the time resolution deteriorates significantly. However, as the pre-shower thickness increases further, the time resolution remains nearly unchanged. This effect may be attributed to the increased number of secondary particles reaching the BGO crystal after the introduction of the pre-shower. Each secondary particle initiates a individual scintillation process, and fluctuations in the start time of these scintillation events contribute to degraded time resolution.

It is worth noting that while some signals with amplitudes below 200 mV originate from particles that bypass the pre-shower and directly hit the test crystal, the final results still reveal a clear distinction between cases with and without a pre-shower. These findings may provide valuable insights for the design of crystal calorimeters, particularly those incorporating longitudinal segmentation.

\section{Conclusion}

This study investigates the time resolution characteristics of various crystal-SiPM detection units under both minimum ionization particle (MIP) and high-energy electron signals. The crystal samples included BGO, PWO, and BSO, with 6, 4, and 1 dimension(s), respectively.

Three timing methods were compared in terms of time resolution. The results indicate that the exponential fitting method combined with approximately 10\% constant fraction timing provides the best timing performance.

Using 10 GeV pion beams, we measured the energy and time responses of different crystal units under a single MIP signal. The results show that time resolution is influenced by both the light yield and waveform rise time. Among crystals of similar lengths, PWO exhibited the best time resolution due to its fast scintillation decay time, leading to a quicker signal rise. BGO and BSO, which have opposite characteristics in light yield and rise time, demonstrated nearly identical time resolutions.

The timing performance of long BGO crystal units with dimensions of 1$\times$1$\times$40 cm$^3$ and 1.5$\times$1.5$\times$60 cm$^3$ was further examined using pion beams. By scanning along the crystal length, we evaluated the uniformity of light yield and time resolution, finding that time resolution remained consistent across different impact positions. Additionally, preliminary particle position localization using the time difference between signals from both ends of the crystal yielded a position resolution of approximately 25 cm for the 40 cm unit and 40 cm for the 60 cm unit.

To study time resolution at different energy levels and electromagnetic shower depths, pre-showers of varying thickness were introduced upstream of the 1$\times$1$\times$40 cm$^3$ and 1.5$\times$1.5$\times$60 cm$^3$ BGO crystal units using electron beams. Time resolution improved with increasing signal amplitude, reaching approximately 200 ps at maximum amplitudes for both crystal units. Additionally, the presence of a pre-shower significantly impacted time resolution. For similar signal amplitudes, time resolution worsened in the presence of a pre-shower, likely due to the increased number of secondary particles absorbed by the crystal, introducing greater time jitter.

Overall, this study provides valuable insights into the time resolution performance of different crystal-SiPM detection units under various conditions. The findings highlight the key factors affecting timing accuracy, including crystal type, length, signal amplitude, and the presence of pre-showers, which are crucial considerations in the design of crystal calorimeters, particularly for applications requiring precise timing measurements.

\section*{Acknowledgments}

The beam test measurements have been performed at the Test Beam Facility at CERN and DESY. The authors sincerely appreciate the technical support and administrative assistance provided by DESY and CERN, as well as the contributions of CALICE in facilitating the successful execution of these beam test experiments. This work was supported by the following funding agencies: National Key R\&D Program of China (Grant No.: 2023YFA1606904 and 2023YFA1606900), National Natural Science Foundation of China (Grant No.: 12150006), Shanghai Pilot Program for Basic Research—Shanghai Jiao Tong University (Grant No.: 21TQ1400209), and National Center for High-Level Talent Training in Mathematics, Physics, Chemistry, and Biology.



\section{Bibliography}

\begin{thebibliography}{99}

\bibitem{CEPCStudyGroup:2023quu}
W.~Abdallah et al. [CEPC Study Group],
CEPC Technical Design Report: Accelerator.
Radiat. Detect. Technol. Methods \textbf{8}, 1-1105 (2024).
doi:10.1007/s41605-024-00463-y, arXiv:2312.14363 [physics.acc-ph].

\bibitem{CEPCCDR-2}
The CEPC Study Group,
CEPC Conceptual Design Report: Volume 2 - Physics \& Detector.
arXiv:1811.10545 [hep-ex] (2018).
URL: \url{https://arxiv.org/abs/1811.10545}.

\bibitem{CERN-LHCC-97-033}
CMS Collaboration,
The CMS electromagnetic calorimeter project: Technical Design Report.
CERN, Geneva (1997).
URL: \url{https://cds.cern.ch/record/349375}.

\bibitem{CMS:2012qbp}
S.~Chatrchyan et al. [CMS Collaboration],
Observation of a New Boson at a Mass of 125 GeV with the CMS Experiment at the LHC.
Phys. Lett. B \textbf{716}, 30-61 (2012).
doi:10.1016/j.physletb.2012.08.021, arXiv:1207.7235 [hep-ex].

\bibitem{SUMNER1988252}
R.~Sumner,
The L3 BGO electromagnetic calorimeter.
Nucl. Instrum. Methods A \textbf{265}, 252-257 (1988).
doi:10.1016/0168-9002(88)91078-9.

\bibitem{ADEVA199035}
B.~Adeva et al.,
The construction of the L3 experiment.
Nucl. Instrum. Methods A \textbf{289}, 35-102 (1990).
doi:10.1016/0168-9002(90)90250-A.

\bibitem{Myers:226776}
S.~Myers,
The LEP Collider, from design to approval and commissioning.
CERN, Geneva (1991).
doi:10.5170/CERN-1991-008.
URL: \url{https://cds.cern.ch/record/226776}.

\bibitem{BaBar:1995bns}
D.~Boutigny et al. [BaBar Collaboration],
BaBar technical design report.
SLAC-R-0457 (1995).

\bibitem{Belle:1995pqe}
M.~T.~Cheng et al. [Belle Collaboration],
A Study of CP violation in B meson decays: Technical design report.
BELLE-TDR-3-95, KEK-95-1 (1995).

\bibitem{abe2010belleiitechnicaldesign}
T.~Abe, I.~Adachi, K.~Adamczyk et al.,
Belle II Technical Design Report.
arXiv:1011.0352 [physics.ins-det] (2010).
URL: \url{https://arxiv.org/abs/1011.0352}.

\bibitem{ABLIKIM2010345}
M.~Ablikim et al.,
Design and construction of the BESIII detector.
Nucl. Instrum. Methods A \textbf{614}, 345-399 (2010).
doi:10.1016/j.nima.2009.12.050.

\bibitem{Liu_2020}
Y.~Liu, J.~Jiang, Y.~Wang,
High-granularity crystal calorimetry: conceptual designs and first studies.
JINST \textbf{15}, C04056 (2020).
doi:10.1088/1748-0221/15/04/C04056.

\bibitem{instruments6030040}
B.~Qi, Y.~Liu,
R\&D of a Novel High Granularity Crystal Electromagnetic Calorimeter.
Instruments \textbf{6}, 40 (2022).
doi:10.3390/instruments6030040.

\bibitem{THOMSON200925}
M.~A.~Thomson,
Particle flow calorimetry and the PandoraPFA algorithm.
Nucl. Instrum. Methods A \textbf{611}, 25-40 (2009).
doi:10.1016/j.nima.2009.09.009.

\bibitem{JI2014143}
Z.~Ji, H.~Ni, L.~Yuan, J.~Chen, S.~Wang,
Investigation of optical transmittance and light response uniformity of 600-mm-long BGO crystals.
Nucl. Instrum. Methods A \textbf{753}, 143-148 (2014).
doi:10.1016/j.nima.2014.03.056.

\bibitem{BACCARO199866}
S.~Baccaro et al.,
Investigation of lead tungstate (PbWO4) crystal properties.
Nucl. Phys. B Proc. Suppl. \textbf{61}, 66-70 (1998).
doi:10.1016/S0920-5632(97)00540-9.

\bibitem{ISHII2002201}
M.~Ishii et al.,
Development of BSO (Bi$_4$Si$_3$O$_{12}$) crystal for radiation detector.
Opt. Mater. \textbf{19}, 201-212 (2002).
doi:10.1016/S0925-3467(01)00220-8.

\bibitem{S13360-6025PE}
HAMAMATSU,
S13360-6025PE.
URL: \url{https://www.hamamatsu.com.cn/cn/zh-cn/product/optical-sensors/mppc/mppc_mppc-array/S13360-6025PE.html}.

\bibitem{986743}
L.~Durieu, M.~Martini, A.-S.~Muller,
Optics studies for the T9 beam line in the CERN PS East Area secondary beam facility.
Proceedings of PAC2001, 1547-1549 (2001).
doi:10.1109/PAC.2001.986743.

\bibitem{DAlessandro:2019ete}
G.~L.~D'Alessandro et al.,
Implementation of CERN secondary beam lines T9 and T10 in BDSIM.
JACoW-IPAC2019-THPGW069 (2019).
doi:10.18429/JACoW-IPAC2019-THPGW069.

\bibitem{Parozzi:2024pfk}
E.~Parozzi et al.,
Secondary beam line efficiency studies at the CERN PS East Experimental Area.
JACoW-IPAC2024-TUPC52 (2024).
doi:10.18429/JACoW-IPAC2024-TUPC52.

\bibitem{DIENER2019265}
R.~Diener et al.,
The DESY II test beam facility.
Nucl. Instrum. Methods A \textbf{922}, 265-286 (2019).
doi:10.1016/j.nima.2018.11.133.

\bibitem{LUCCHINI2023168214}
M.~T.~Lucchini et al.,
Sub-10 ps time tagging of electromagnetic showers with scintillating glasses and SiPMs.
Nucl. Instrum. Methods A \textbf{1051}, 168214 (2023).
doi:10.1016/j.nima.2023.168214.

\end{thebibliography}

\end{document}